\documentclass[prc,reprint,amsmath,amssymb,aps,nofootinbib]{revtex4-2}
\usepackage{subfigure}
\usepackage{graphicx}
\usepackage{bm}
\usepackage{amsfonts}
\usepackage{siunitx}
\graphicspath{{./figs/}{./}}

\newcommand{\com}[1]{{#1}}

\newcommand{\im}{{\rm Im}}
\newcommand{\re}{{\rm Re}}

\newcommand{\fm}{\si{\femto\meter}}

\newcommand{\mev}{\si{\mega\electronvolt}}
\newcommand{\gev}{\si{\giga\electronvolt}}

\newcommand{\mG}{\mathcal{G}}

\newcommand{\mO}{\mathcal{O}}
\newcommand{\mS}{\mathcal{S}}

\newcommand{\rcnp}{\affiliation{Research Center for Nuclear Physics (RCNP), Osaka University, Ibaraki, Osaka 567-0047, Japan}}
\newcommand{\titech}{\affiliation{Department of Physics, Tokyo Institute of Technology, Meguro, Tokyo 152-8551, Japan}}
\allowdisplaybreaks[1]
\newcommand{\bp}{\bm{p}}

\newcommand{\bl}{\bm{l}}

\begin{document}
\title{Spectral function of the $\eta'$ meson in nuclear medium based on phenomenological models}
\date{\today}

\author{Shuntaro Sakai}
\email{shsakai@rcnp.osaka-u.ac.jp}
\rcnp

\author{Daisuke Jido}
\titech

\begin{abstract} 
The in-medium modification of the spectral function of the $\eta'$ meson with and without the spatial momentum is studied with the $T\rho$ approximation by employing 
two phenomenological models for the $\eta'N$ scattering;
one is called \com{coupled channels model} and the other the $N(1895)$-dominance model.
 In the former model, the $\eta'N$ scattering amplitude is calculated in the unitarized coupled-channel approach involving the $\eta'N$ channel, while in the latter model the $\eta'N$ scattering process is dominated by the $N(1895)$ resonance with the spin and parity $J^P=1/2^-$.
 In the coupled channels model, one single peak of the in-medium $\eta'$ mode appears in the spectral function and the peak position shifts to higher energies along with the increase of the nuclear density reflecting the repulsive $\eta'N$ scattering length of the unitarized coupled-channel amplitude.
On the other hand, two branches related to the $\eta'$ and $N(1895)$-hole modes appear in the $N(1895)$-dominance model.
In both models, the shift of the peak position and the width in the spectral function are a few tens of $\mev$ at the normal nuclear density for the $\eta'$ meson at rest in the nuclear medium.
Once the spatial momentum is turned on, the peak positions in the spectral function approach the energies without the nuclear medium effect.
Particularly, in the $N(1895)$-dominance model, the peak strength of the $N(1895)$-hole mode gets smaller with the finite momentum and the spectral function comes to have one single peak.
\end{abstract}
\maketitle

\section{Introduction}
The study of the $\eta'$ meson in the nuclear medium is one of the most interesting topics in hadron physics, and it provides clues to understand the properties of chiral 
symmetry and axial $U(1)$ symmetry in the nuclear medium.
The mass of the $\eta'$ meson is raised by the $U_A(1)$ anomaly, which explicitly breaks the $U_A(1)$ symmetry 
and is tied to the nonperturbative gluon dynamics such as instanton or topological structure of the QCD vacuum (see, e.g., Refs.~\cite{Christos:1984tu,tHooft:1986ooh} for the review articles).
Meanwhile, as pointed out in Refs.~\cite{Lee:1996zy,Jido:2011pq}, the breaking of chiral symmetry is indispensable for the $U_A(1)$ anomaly to come into play in the meson mass spectrum when the number of the flavor is equal to or larger than three.
Many calculations have been done with various models to study the mass shift of the $\eta'$ meson in the nuclear medium, for instance, in Refs.~\cite{Pisarski:1983ms,Bernard:1987sg,Bernard:1987sx,Kapusta:1995ww,Costa:2002gk,Nagahiro:2006dr,Jido:2011pq,Nagahiro:2011fi,Sakai:2013nba,Fejos:2017kpq,Suenaga:2019urn}, and the possible formation of the $\eta'$-nucleus bound state is discussed in Refs.~\cite{Tsushima:1998qw,Nagahiro:2004qz,Bass:2005hn,Nagahiro:2004qz,Saito:2005rv,Itahashi:2012ut,Jido:2018aew}.
It should be noted that Refs.~\cite{Jido:2011pq,Sakai:2013nba} have suggested that the $\eta'$ mass be reduced in the nuclear medium, where chiral symmetry is partially restored, as a consequence of the reduction of the magnitude of the quark condensate irrelevantly to the fate of the $U_{A}$(1) anomaly in the nuclear medium. This $\eta'$ mass reduction in the nuclear matter 
should be
observed in finite nuclei as an attractive scalar potential for the $\eta'$ meson in nuclei. With this attractive potential one may expect some nuclear bound states of the $\eta'$ meson. 
Motivated by the above-mentioned studies,
some experiments to clarify the in-medium $\eta'$ properties have been carried out~\cite{CBELSATAPS:2016qdi,Friedrich:2016cms,n-PRiMESuper-FRS:2016vbn,e-PRiMESuper-FRS:2017bzq,CBELSATAPS:2018sck,LEPS2BGOegg:2020cth};
particularly as reported in Refs.~\cite{n-PRiMESuper-FRS:2016vbn,e-PRiMESuper-FRS:2017bzq,LEPS2BGOegg:2020cth}, no signal of the $\eta'$-nucleus bound state was observed so far.
We refer to the review articles Refs.~\cite{Bass:2018xmz,Bass:2021rch} on the physics of the $\eta'$ meson and the $\eta'$-nucleus system.

One notable point is that systems in the nuclear medium are not Lorentz invariant and the $\eta'$
meson dispersion relation can be modified.
Then, the peak position of the invariant mass observed by the decay products does not correspond to the in-medium mass of the particle directly when the spatial momentum of the decaying particle is finite.
Therefore, this effect needs to be taken into consideration in order to reach better understanding or interpretation of the experimental data.
The spectral function of the light mesons in the nuclear medium with finite momentum is studied in Refs.~\cite{Lee:1997zta,Friman:1997tc,Peters:1997va,Saito:1998wd,Post:2003hu,Cabrera:2009ep,Kim:2019ybi}.
Particularly, the vector-meson properties including the finite-momentum effect have been studied intensively in the theoretical side as in the aforementioned works, and regarding the $\phi$ meson an experiment at J-PARC is forthcoming for the systematic study of the in-medium properties such as the spectral function and dispersion relation with better statistics and resolution compared with those achieved in the previous experiment~\cite{KEK-PS-E325:2005wbm} (see, e.g., Refs.~\cite{proporsalE16,Aoki:2015qla} for the details of the planned experiment).
One of the recent theoretical developments on the spectral function of the vector mesons is to apply the functional renormalization group such as in Ref.~\cite{Tripolt:2021jtp}.

In this work, we investigate the in-medium spectral function of the $\eta'$ meson with finite spatial momentum. 
As the nuclear medium is composed of nucleons, the investigation of the in-medium $\eta'$ properties based on the two-body scattering of $\eta'$ and nucleon is a first promising way particularly in low densities.
Here, the driving force of the in-medium $\eta'$ modification is the $\eta'N$ scattering.
We employ two models for possible scenarios of the $\eta'N$ scattering amplitude;
one is the \com{coupled channels model} and the other is the $N(1895)$-dominance model.
In the \com{coupled channels model}, the scattering amplitude is constructed on the basis of the meson-baryon dynamics and the scattering process is driven by the interaction kernel without explicit resonance degree of freedom. 
Some theoretical works are made for the construction and the application of the chiral effective model containing $\eta'$ and nucleon~\cite{Kawarabayashi:1980uh,Borasoy:1999nd,Bass:1999is}, and the $\eta'N$ interaction is studied in Refs.~\cite{Borasoy:2002mt,Bass:2005hn,Oset:2010ub,Sakai:2013nba,Sakai:2016vcl,Gao:2017hya,Bruns:2019fwi}.
On the other hand, in the $N(1895)$-dominance model, the $\eta'N$ scattering amplitude is obtained by the process through the $N(1895)$ resonance.
The existence of the $N(1895)$ resonance with $J^P=1/2^-$ near the $\eta'N$ threshold has been suggested by the recent analyses of the experimental data including the $\eta$ and $\eta'$ photoproduction process~\cite{Anisovich:2017pox,A2:2017gwp,Anisovich:2018yoo,Tiator:2018heh,CBELSATAPS:2020cwk},
and this resonance is now listed as a four-star state in the recent version of the Review of Particle Physics (RPP)~\cite{ParticleDataGroup:2022pth}.
The $N(1895)$ resonance has also been studied from theoretical view points in Refs.~\cite{Khemchandani:2013nma,An:2018vmk,Khemchandani:2020exc}.
The presence of the near-threshold resonance which couples to the $s$-wave $\eta'N$ pair can have a large impact on the in-medium $\eta'$ properties through the $N(1895)$-hole excitation.
The details of the model will be explained in the following section.
An experiment for the study of the in-medium $\eta'$ spectral function is in progress by the LEPS2 Collaboration at SPring-8,
and we hope that theoretical works related to the in-medium properties of the $\eta'$ meson give some clues and indication for the understanding and interpretation of the data expected in the future.

The paper is organized as follows. In Sec.~\ref{sec:preliminaries} we define the in-medium quantities of the $\eta'$ meson and show the $\eta'$ spectral function by using 
an $\eta'N$ scattering amplitude of the effective range expansion in the $T\rho$ approximation in order to see how the spectral function is modified in the nuclear medium. In Sec.~\ref{sec:models}, we explain the models which we use in this work. In Sec.~\ref{sec:results}, we show our numerical results, and Sec.~\ref{sec:summary} is devoted to summary and conclusion.

\section{Preliminaries}
\label{sec:preliminaries}
First of all, we define the in-medium quantities of the $\eta'$ meson by using its propagator with the self-energy. We consider the spin and isospin symmetric uniform
nuclear matter here for the isospin singlet pseudoscalar $\eta'$ meson.
With a given in-medium self energy of the $\eta'$ meson, $\Pi_{\eta'}$, the $\eta'$ propagator in the nuclear medium, $D_{\eta'}$, is obtained by
\begin{align}
 D_{\eta'}(\omega,\bp;\rho)=&\frac{1}{\omega^2-\bp^2-m_{\eta'}^2-\Pi_{\eta'}(\omega,\bp;\rho)+i\epsilon},\label{eq:spec1}
\end{align}
where $\rho$ is the nuclear density, $m_{\eta'}$ is the $\eta'$ meson mass in vacuum,
and the four momentum of the $\eta'$ meson is set as $p_{\eta'}^\mu=(\omega,\bp)$ in the nuclear medium rest frame.
With the in-medium propagator $D_{\eta'}$, the $\eta'$ spectral function in the nuclear medium is written as
\begin{align}
 \mS_{\eta'}=&-\frac{1}{\pi}\im\left(D_{\eta'}\right).\label{eq:spec}
\end{align}
The in-medium properties of the $\eta'$ meson can be read from the in-medium propagator $D_{\eta'}$ at the pole. 
The pole position of $D_{\eta'}$ for $\eta'$ at rest $\bp=\bm{0}$, $\omega_P$, is obtained by 
\begin{align}
 D_{\eta'}^{-1}(\omega_P,\bm{0};\rho)=\omega_P^2-m_{\eta'}^2-\Pi_{\eta'}(\omega_P,\bm{0};\rho)=0,\label{eq:propinv}
\end{align}
{and $\omega_{P}$ corresponds to the rest mass of the in-medium $\eta'$ meson.}   
We parameterize 
$\omega_P$ as 
\begin{align}
\omega_P^2=\omega_R^2-i\omega_R \Gamma_*,  \label{eq:mass}
\end{align}
with the in-medium mass $\omega_R$ and width $\Gamma_*$\footnote{One may define the in-medium mass and width as the pole position of the in-medium propagator in the complex $\omega$ plane. In this case the pole position is parameterized as $\omega_P=\omega_R - i \Gamma_{*}/2$. In the present calculation, these two definitions provide just a slight difference.}, and these quantities are given by
\begin{align}
 \begin{split}
   \omega_R^{2} =& m_{\eta'}^{2} + \re\left(\Pi_{\eta'}(\omega_P,\bm{0};\rho)\right), \\
   \Gamma_* =& - \frac{1}{\omega_R} \im\left(\Pi_{\eta'}(\omega_P,\bm{0};\rho)\right) .
 \end{split}\label{eq:MassWidth}
\end{align}
With the pole position $\omega_P$ the in-medium propagator can be written in the relativistic Breit-Wigner form as
\begin{align}
 D_{\eta'}(\omega,\bm{0};\rho)=\frac{Z(\omega)}{\omega^2-\omega_R^2+i\omega_R\Gamma_*}, \label{eq:spec3}
\end{align}
where $Z(\omega)$ in the numerator is the residue function and $Z \equiv Z(\omega_P)$ gives the wave function renormalization at the pole.
The in-medium width $\Gamma_*$ represents the nuclear absorption of the $\eta'$ meson. 

For the $\eta'$ meson moving in the nuclear medium with momentum~$\bp$, 
the pole position of the propagator $D_{\eta'}$ depends on the momentum and $\omega_P(\bp)$ is evaluated by
\begin{align}
\lefteqn{ D_{\eta'}^{-1}(\omega_P(\bp),\bp; \rho)} \\ & =
\omega_P^2(\bp)-\bp^2-m_{\eta'}^2-\Pi_{\eta'}(\omega_P(\bp),\bp;\rho)=0.\label{eq:propinv2}
\end{align} 
The pole position $\omega_P(\bp)$ gives the dispersion relation of the $\eta'$ meson in the nuclear medium.
Owing to the breaking of the Lorentz invariance by the presence of the nuclear medium, the invariant mass squared at the pole position, $p_{\eta'}^{2}=\omega_{P}^{2}(\bp)-\bp^{2}$, is not necessarily equal to the rest mass squared of the in-medium $\eta'$ meson:
\begin{align}
   p_{\eta'}^{2}= \omega_P^2 + \left( \Pi_{\eta'}(\omega_P(\bp),\bp;\rho) - \Pi_{\eta'}(\omega_P(\bm 0),\bm 0;\rho)\right),  \label{eq:invmass}
\end{align}
where we have used Eqs.~\eqref{eq:propinv2} and \eqref{eq:propinv}. If the self-energy is Lorentz-invariant, the second term of the right hand side of Eq.~\eqref{eq:invmass} vanishes and the invariant mass of the $\eta'$ meson $p_{\eta'}^{2}$ is also Lorentz-invariant. But in the presence of the nuclear medium the Lorentz invariance can be broken, the second term does not vanish any more. 
In such a case, it is important to note that the peak position of $\mS_{\eta'}$ as a function of the $\eta'$ invariant mass $\sqrt{p_{\eta'}^2}=(\omega^2-\bp^2)^{1/2}$ does not necessarily correspond to the in-medium $\eta'$ mass.

Expanding $D_{\eta'}^{-1}$ around $\omega^2=\omega_P^2$ and $\bp^2= 0$, we identify the wave function 
renormalization $Z$ and 
the velocity $\beta$ as Ref.~\cite{Goda:2013npa} by
\begin{align}
 D_{\eta'}^{-1}(\omega,\bp; \rho)=&Z^{-1}\left[\omega^2-\beta^2\bp^2-\omega_P^2\right]+\cdots .\label{eq:spec2}
\end{align}
Comparing Eq.~\eqref{eq:spec2} with Eq.~\eqref{eq:spec1}, we find $Z$ and $\beta^2$ written with the in-medium self energy as
\begin{align}
 Z&=\left.\left(1-\frac{\partial \Pi_{\eta'}}{\partial\omega^2}\right)^{-1}\right|_{\omega^2=\omega_P^2,\bp^2=0},\\
 \beta^2&=\left.Z\left(1+\frac{\partial \Pi_{\eta'}}{\partial\bp^2}\right)\right|_{\omega^2=\omega_P^2,\bp^2=0}.
\end{align}
These quantities $Z$ and $\beta$ defined at the pole in the complex energy plane $\omega^2=\omega_P^2$ can be complex numbers in general.
The velocity $\beta$ describes the dispersion relation with a small spatial momentum.
Due to the breaking of the Lorentz invariance, the velocity $\beta$ is allowed to take a different value from unity. With the deviation of the velocity from unity, the invariant mass squared at the pole for small spatial momenta is written as $p_{\eta'}^2=\omega_P^2+(\beta^2-1)\bp^2$ and is to be spatial-momentum dependent~\cite{Lee:1997zta}.

As seen in Eq.~\eqref{eq:spec1}, the nuclear medium effect is contained in the in-medium self energy~$\Pi_{\eta'}$.
The interaction of the $\eta'$ meson in the nuclear medium is not known yet.
In this work, we evaluate the in-medium $\eta'$ self energy  
using relatively better known $\eta'N$ scattering. In this work, we focus on the one-nucleon processes.\footnote{
The two-body effects are found to be smaller compared with the one-body effects in Ref.~\cite{Nagahiro:2011fi}.}
The $\eta'$ meson is scattered by a nucleon bound in the nuclear medium where
the nucleons fill the Fermi sphere up to the Fermi momentum $k_f=(3\pi^2\rho/2)^{1/3}$.
{We calculate the $\eta'$ self energy by using the $s$ wave $\eta'N \to \eta'N$ scattering $T$ matrix $T_{\eta'N}$ as}
\begin{align}
 \Pi_{\eta'}=&4\int\frac{d^3l}{(2\pi)^3}
 T_{\eta'N}(\sqrt{s})\theta(k_f-l),\label{eq:se_tmp}
\end{align}
where the $\eta'N$ invariant mass squared $s=(\omega+E_N)^2-|\bp+\bl|^2$ is evaluated by the nucleon four momentum $p_N^\mu=(E_N,\bl)$, and
the factor $4$ accounts for the spin and isospin degeneracy, and the step function $\theta(k_f-l)$ represents the nucleon occupation number in the Fermi gas approximation.
One may consider the on-shell $T$-matrix in the integral. In such a case,
the invariant mass $\sqrt s$ of the $T$-matrix is fixed by the external energy 
and $T_{\eta'N}$ in the integrand does not depend on the nucleon momentum $l$. 
Then, the $\eta'N$ scattering $T$ matrix $T_{\eta'N}$ can be factored out from the integral, and
the integral over $l$ just gives the nuclear density $\rho$.
In this way, Eq.~\eqref{eq:se_tmp} is reduced to the following expression of the self energy with the $T\rho$ approximation in the end;
\begin{align}
 \Pi_{\eta'}(\omega,\bp;\rho)=T_{\eta'N}(\sqrt{s})\rho,\label{eq:trho2}
\end{align}
where the $\eta'N$ invariant mass $\sqrt s$ is evaluated as $s=(\omega+m_N)^2-\bp^2$ where the nucleon is at rest $p_N^\mu=(m_N,\bm{0})$.
With the $\eta'N$ scattering amplitude, which will be evaluated with some certain models, we obtain the spectral function, mass, width, velocity, and wave function renormalization of the $\eta'$ meson in nuclear medium.
In this prescription, the nucleon is at rest in the nuclear medium and its mass is identical to the one at $\rho=0$.
The nucleon is actually bound in nuclear matter, and the Pauli-blocking and nuclear binding effect can enter the nucleon energy and 
the scattering process~\cite{Waas:1996xh,Waas:1996fy,Waas:1997pe,Ramos:1999ku,Sekihara:2012wj}.
For example, in Refs.~\cite{Waas:1996xh,Sekihara:2012wj}, the nucleon energy is reduced by $\mO(k_f^2/m_N^2)\sim 10\%$ at the normal nuclear density due to the nuclear binding effect.
Here, we accept this $10\%$ as a typical order of the uncertainties originating from the treatment of the nuclear medium effect and focus on the qualitative feature of the spectral function and pole structure obtained with the $T\rho$ approximation.

Before moving to the details of the models, we demonstrate
how the spectral function of the $\eta'$ meson
is modified with the nuclear medium effects
on the $\eta'$ meson by employing simple models for the $\eta'N$ scattering amplitude. 
First of all, we consider the scattering-length approximation.
In this approximation, the $\eta'N$ scattering $T$ matrix $T_{\eta'N}(\sqrt{s})$ in Eq.~\eqref{eq:trho2} is treated as an energy-independent constant evaluated at the threshold:
\begin{align}
 T_{\eta'N}(\sqrt{s}=m_N+m_{\eta'})=-\frac{8\pi(m_N+m_{\eta'})}{2m_N}a_{\eta'N},\label{eq:rel_sl_tamp}
\end{align}
where $a_{\eta'N}$ is the $\eta'N$ scattering length.
With the $T$-matrix~\eqref{eq:rel_sl_tamp}, the inverse of the in-medium $\eta'$ propagator $D_{\eta'}$ is written as
\begin{align}
\hspace{-20pt}
 D_{\eta'}^{-1}=\omega^2-\bp^2-m_{\eta'}^2-\left(-\frac{8\pi(m_N+m_{\eta'})}{2m_N}\right)a_{\eta'N}\rho,\label{eq:d_slapprx}
\end{align}
and the pole position is found as
\begin{align}
 \omega_P^{2}(\bp) =m_{\eta'}^2-4\pi\left(1+\frac{m_{\eta'}}{m_N}\right)a_{\eta'N}\rho + \bp^{2}.
\end{align}
Thus, one can easily see that the positive (negative) $a_{\eta'N}$, which corresponds to the attractive (repulsive) sign, leads to the mass reduction (increase) in the nuclear medium.
In spite of the inclusion of the medium effect on the $\eta'$ meson,
since neither of energy nor momentum dependences is considered in this approximation,
the pole position does not have a nontrivial momentum dependence. 
Thus, the dispersion relation $\omega_P(\bp)$ has the same momentum dependence with the free-space one. 

\begin{figure}[t]
 \centering
 \includegraphics[width=0.45\textwidth]{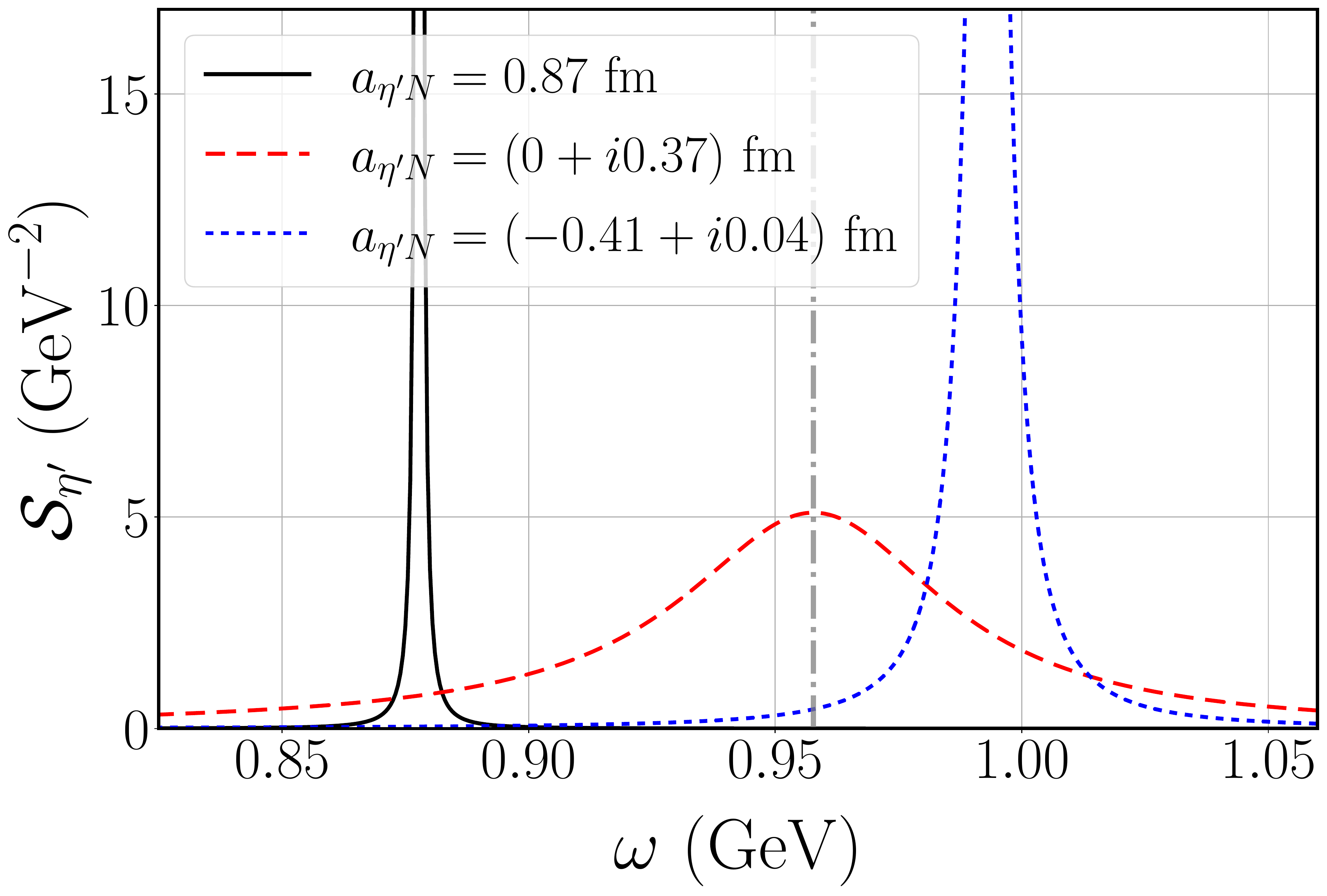}
 \caption{In-medium $\eta'$ spectral functions $\mS_{\eta'}$ calculated with the in-medium $\eta'$ propagator~\eqref{eq:d_slapprx} as functions of the $\eta'$ energy $\omega$
for the $\eta'N$ scattering length $a_{\eta'N}=0.87~\fm$ (black-solid), $(0+i0.37)~\fm$ (red-dashed), and $(-0.41+i0.04)~\fm$ (blue-dotted).
The vertical dashed-dotted line denotes the in-vacuum $\eta'$ mass.
}
 \label{fig:spec_mom_dem_2a}
\end{figure}

In Fig.~\ref{fig:spec_mom_dem_2a}, we show the spectral functions calculated with the in-medium propagator~\eqref{eq:d_slapprx} 
{for the $\eta'$ meson at rest, $\bm p = \bm 0$, at the normal nuclear medium in the scattering length approximation. Here we consider three examples for the $\eta'N$ scattering length, $a_{\eta'N} =+0.87~\fm$, $a_{\eta'N}=(0+i0.37)~\fm$ and $(-0.41+i0.04)~\fm$. The first case is a theoretical estimation corresponding to the scattering length which provides $80~\mev$ mass reduction at the normal density $\rho=\rho_0$. This mass reduction is obtained by a linear $\sigma$ model~\cite{Sakai:2013nba}, and similar mass reductions are also suggested by other model calculations~\cite{Costa:2002gk,Nagahiro:2006dr,Suenaga:2019urn}. The second case is the central value of $a_{\eta'N} = 0^{+0.43}_{-0.43} + i 0.37^{+0.40}_{-0.16}~\fm$ which is experimentally extracted from the low-energy $pp\to pp\eta'$ data~\cite{Czerwinski:2014yot}.\footnote{The scattering length given by the partial wave analysis~\cite{Anisovich:2018yoo} is consistent with Ref.~\cite{Czerwinski:2014yot}.} The third one is a phenomenological evaluation by  a hadronic scattering model~\cite{Bruns:2019fwi}. This phenomenological model will be explained in detail later as coupled channels model.
}
As shown in Fig.~\ref{fig:spec_mom_dem_2a}, 
the peak position of the spectral function $\mS_{\eta'}$ with $a_{\eta'N}=0.87~\fm$ is located around $\omega= 0.88~\gev$
which is $80~\mev$ below the in-vacuum $\eta'$ mass $m_{\eta'}=0.958~\gev$.
The peak in the spectral function with $a_{\eta'N}=(0+i0.37)~\fm$ is just broadened by the imaginary part of the scattering length
and no shift of the peak position from the in-vacuum $\eta'$ mass is observed due to the absence of the real part in the scattering length.
The peak position of the spectral function with $a_{\eta'N}=(-0.41+i0.04)~\fm$ gets higher than the $\eta'$ mass at $\rho=0$ reflecting the repulsive sign of the real part of the scattering length.
This peak is relatively narrow as the imaginary part of the scattering length is small.
Thus, qualitatively different spectral functions $\mS_{\eta'}$ can be obtained with these scattering lengths.

\begin{figure*}[t]
 \centering
 \includegraphics[width=\textwidth]{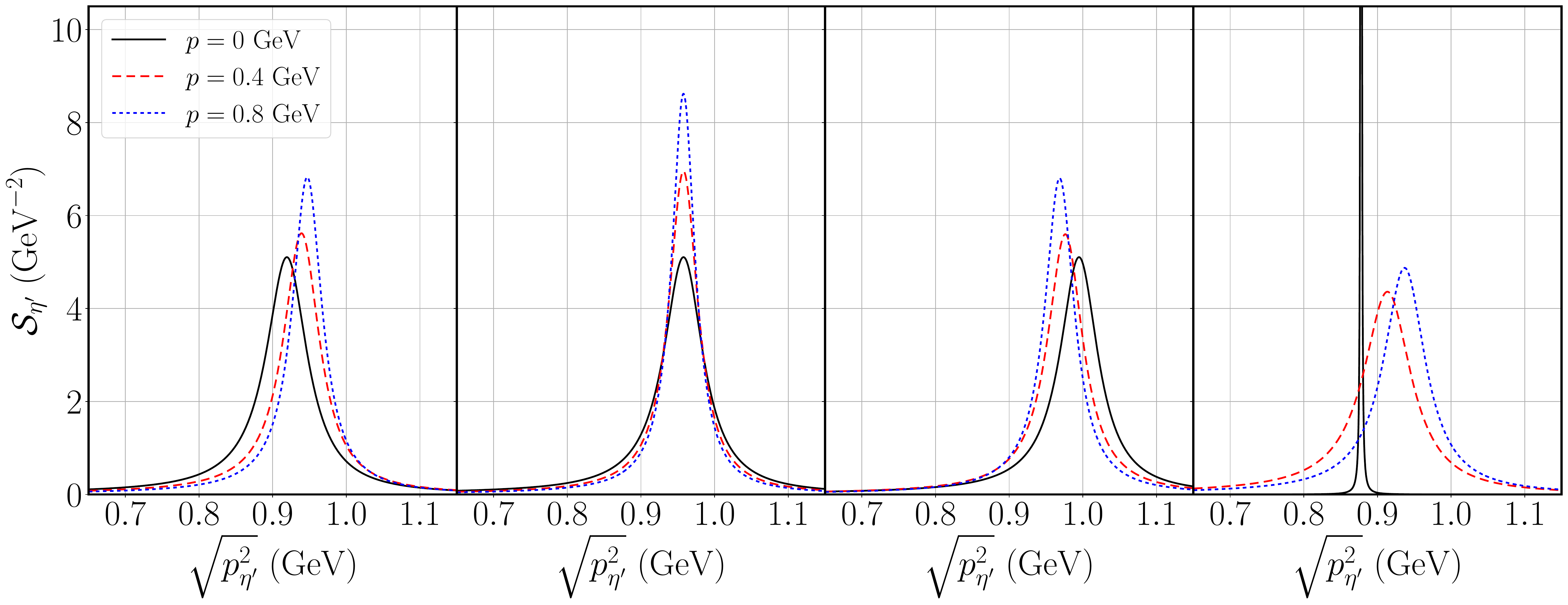}
 \caption{Spectral functions at $\rho=\rho_0$ with fixed values of the $\eta'$ momentum, $p = 0$, $0.4$ and $0.8~\gev$ as functions of the $\eta'$ invariant mass $\sqrt{p_{\eta'}^2}=(\omega^2-\bp^2)^{1/2}$.
 The scattering length in the $T$-matrix is fixed as 
{$a_{\eta'N}=(+0.43+i0.37)~\fm$, $a_{\eta'N}=(0+i0.37)~\fm$, $a_{\eta'N}=(-0.43+i0.37)~\fm$ and $a_{\eta'N} = +0.87~\fm$}
 from the left plot.}
 \label{fig:spec_mom_dem_3}
\end{figure*}

To exhibit the modification of
the in-medium dispersion relation of the $\eta'$ meson originating from the momentum dependence of the $\eta'N$ amplitude,
we consider an $\eta'N$ scattering $T$ matrix in a form of the effective range expansion given by
\begin{align}
 T_{\eta'N}(\sqrt{s})=-\frac{8\pi\sqrt{{s}}}{2m_N}\frac{1}{1/a_{\eta'N}-ip'},
\label{eq:tslmod}
\end{align}
where the variable $\sqrt s$ and $p'$ are the energy and the magnitude of the $\eta'$ momentum in the $\eta'N$ center of mass frame, respectively, and $p'$ is calculated by $p'=\lambda^{1/2}(s,p_{\eta'}^2,m_N^2)/(2\sqrt{{s}})$ with $\lambda(x,y,z)=x^2+y^2+z^2-2(xy+yz+zx)$. 
The momentum $p'$ is generally complex in the complex energy plane and purely imaginary for the real energy below the $\eta'N$ threshold. Here, we just consider the linear $p'$ term in the denominator, which is required by the elastic unitarity, for simplicity.
It should be noted that this prescription to include the momentum dependence in the $\eta'N$ scattering amplitude is minimal, and one may add a $p'^2$ term in the denominator for further momentum dependence in the effective range expansion.
In Fig.~\ref{fig:spec_mom_dem_3} we show the spectral functions $\mS_{\eta'}$ calculated with the $T$-matrix~\eqref{eq:tslmod} for three fixed $\eta'$ spatial momenta $p\equiv|\bp| = 0,\ 0.4,\ 0.8\ \gev$.
The spectral functions are plotted as functions of the invariant mass of the $\eta'$ meson, $\sqrt{p_{\eta'}^2}=(\omega^2-\bp^2)^{1/2}$. 
Here we consider four values of the scattering length;
{$a_{\eta'N}=(0+i0.37)~\fm$, $a_{\eta'N}=(\pm 0.43+i0.37)~\fm$ and $a_{\eta'N}=+0.87~\fm$. The real parts of $a_{\eta'N}=(\pm 0.43+i0.37)~\fm$ correspond to the upper and lower bounds of the experimental uncertainties in Ref.~\cite{Czerwinski:2014yot}.}
For $a_{\eta'N}=(0+i0.37)~\fm$, the $\eta'$ self energy is pure imaginary, and the significant modification of the spectral function appears only in the width of the peak structure. 
The peak of the spectral function $\mS_{\eta'}$ gets sharper with larger $p$ without shift of the peak position.
On the other hand, 
{for $a_{\eta'N}=(\pm 0.43+i0.37)~\fm$,}
the peak position moves to higher (lower) invariant masses with the increase of momentum $p$.
The momentum in the $\eta'N$ c.m.\ frame $p'$ in $T_{\eta'N}(\sqrt{{s}})$ becomes larger with the increase of the $\eta'$ momentum $p$ in the nucleon rest frame.
Then, the real part of the in-medium $\eta'$ self energy $\Pi_{\eta'}$ gets smaller with larger $p$ and the peak approaches the in-vacuum $\eta'$ mass.
This implies that the nuclear medium effect on the in-medium mass shift gets less noticeable for larger spatial momentum of the $\eta'$ meson.
{For $a_{\eta'N}=+0.87~\fm$, because the scattering length has no imaginary part, the peak of the spectral function has no width for $p=0$. Nevertheless, the momentum dependence of the scattering matrix introduces imaginary part to the self-energy, and as a consequence the peaks of the spectral functions for the finite spatial momenta have widths.}
In terms of the velocity of the $\eta'$ meson in the nuclear medium, the real part of the velocity gets $\re(\beta)>1$ for $\re(a_{\eta'N})>0$. In this case the peak position of $\mS_{\eta'}$ shifts to larger $\sqrt{p_{\eta'}^2}$.
The opposite behavior takes place for $\re(a_{\eta'N})<0$.

\begin{figure*}
 \centering
 \subfigure[]{\includegraphics[width=5.7cm]{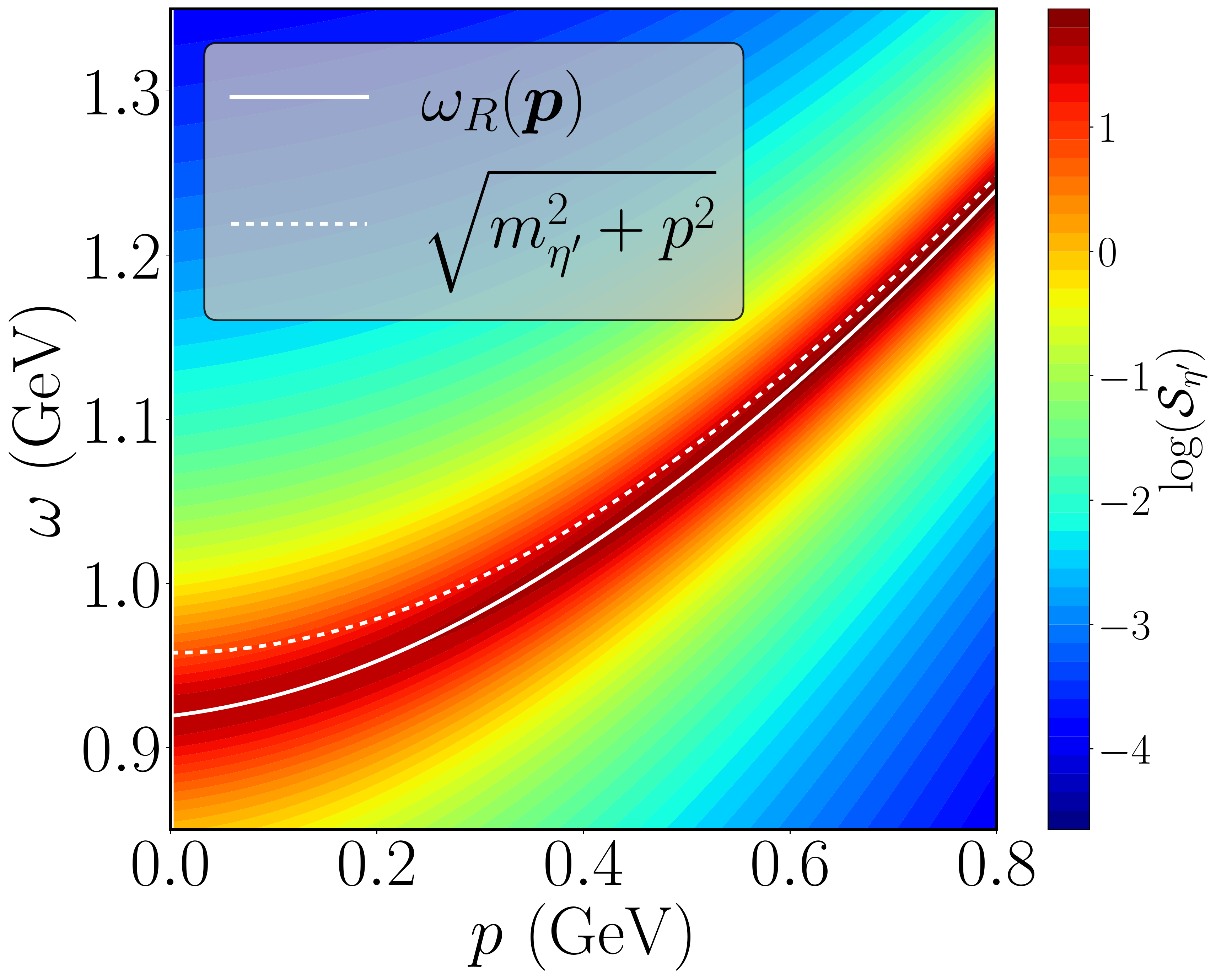}\label{fig:spec_pomega_1a}}\quad
 \subfigure[]{\includegraphics[width=5.7cm]{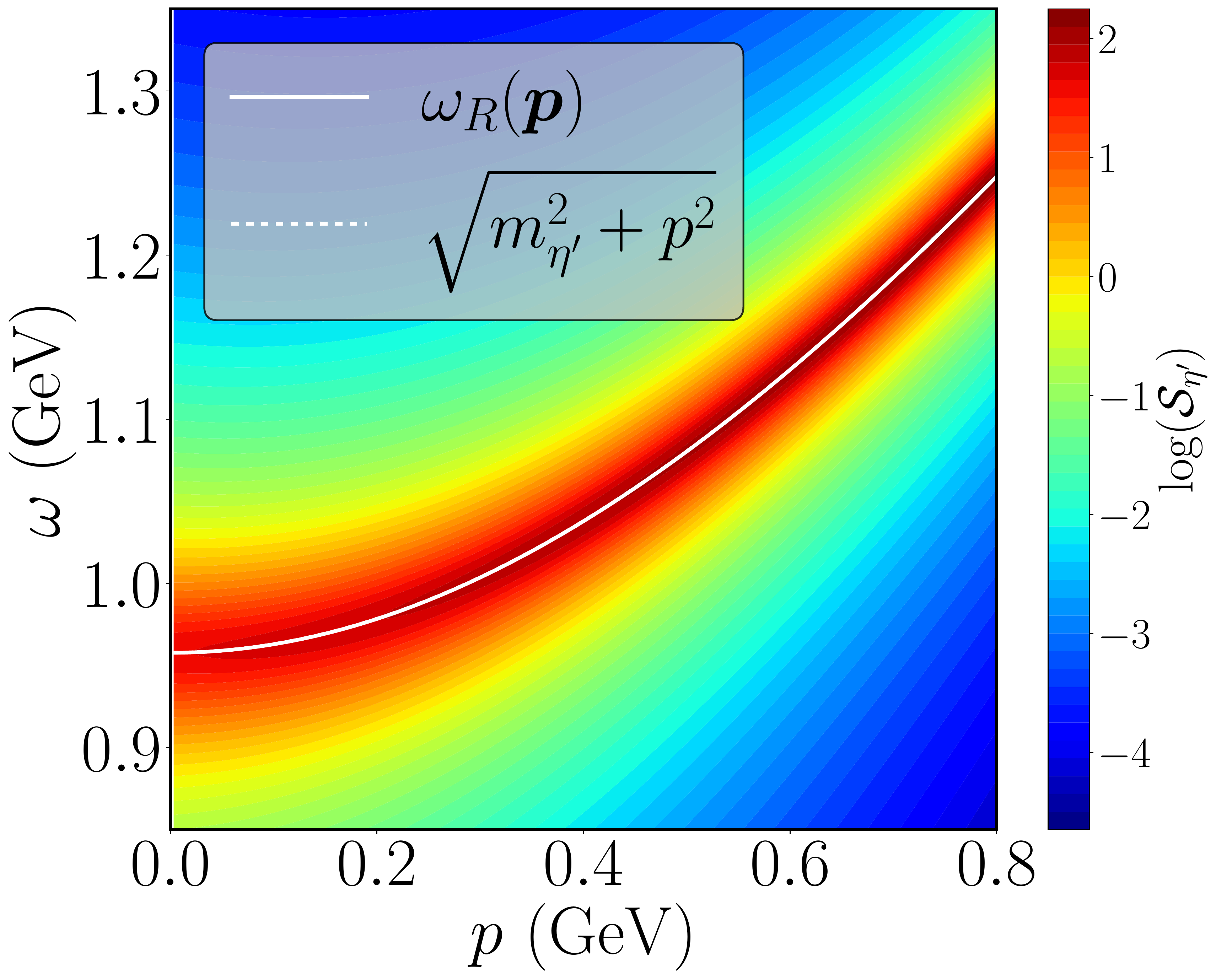}\label{fig:spec_pomega_1b}}\quad
 \subfigure[]{\includegraphics[width=5.7cm]{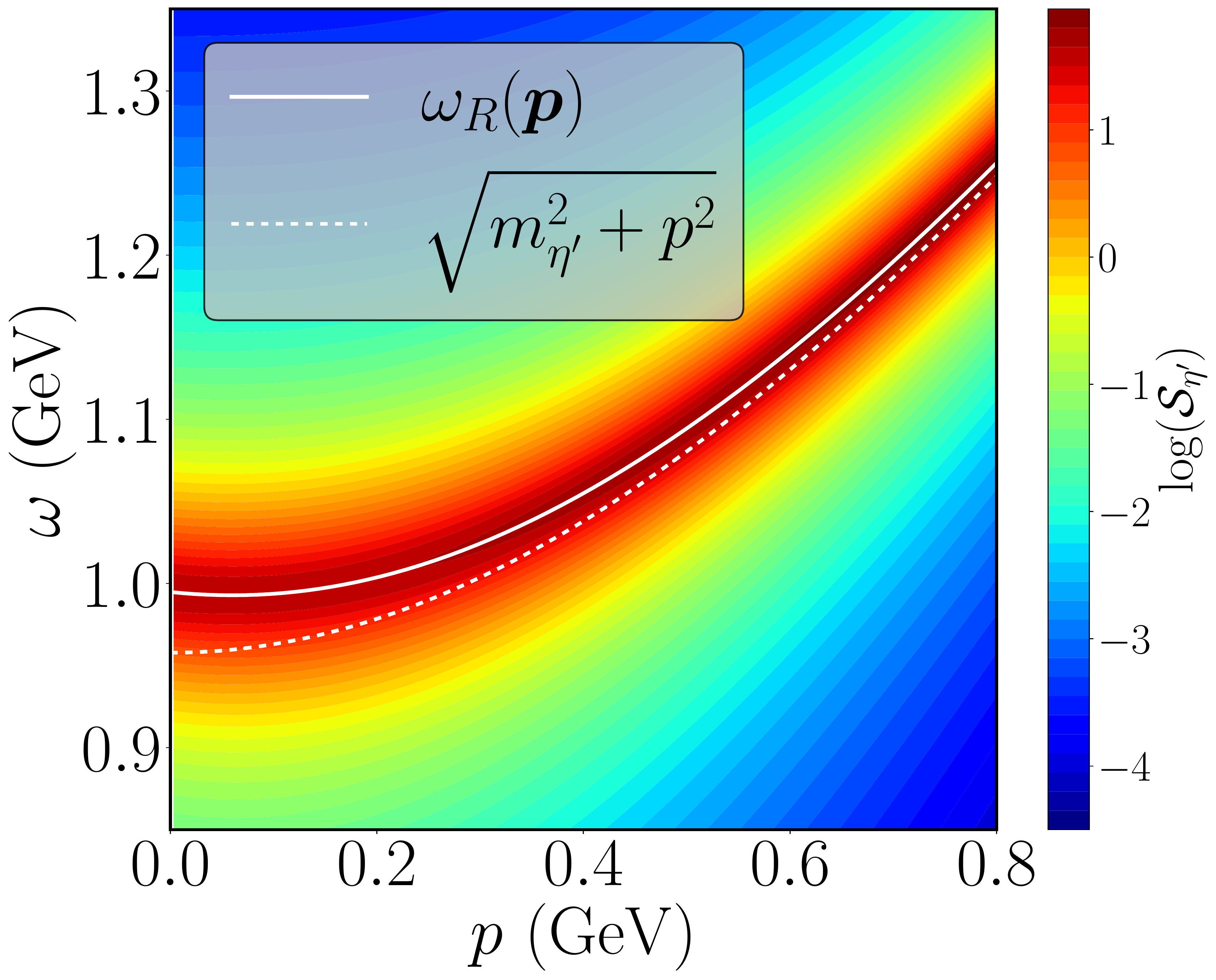}\label{fig:spec_pomega_1c}}
 \caption{ Contour plots of the logarithm of the $\eta'$ spectral function in the $p$-$\omega$ plane for $\rho=\rho_0$. The scattering length in the $T$-matrix is fixed as~$a_{\eta'N}=(+0.43+i0.37)~\fm$ (left), $a_{\eta'N}=(0+i0.37)~\fm$ (middle) and $a_{\eta'N}=(-0.43+i0.37)~\fm$ (right).
 The solid lines are the plots of $\omega_R(\bm{p})$ as functions of $p$ and the dotted lines are the $\eta'$ dispersions in vacuum.
 In the middle panel, the solid and dotted lines are overlapping with each other.}
 \label{fig:spec_pomega}
\end{figure*}

We show in Fig.~\ref{fig:spec_pomega} the contour plots of the logarithm of the spectral function, $\log\left(\mS_{\eta'}\right)$, in the $p$-$\omega$ plane at $\rho=\rho_0$ for the scattering lengths, $a_{\eta'N}=(0+i0.37)~\fm$ and $a_{\eta'N}=(\pm 0.43+i0.37)~\fm$.
In the figure, the solid lines denote the momentum dependence of the real part of the pole position, $\omega_R(\bp)$, 
which is obtained from $\omega_P^2(\bp)=\omega_R^2(\bp)-i\omega_R(\bp)\Gamma_*(\bp)$ for the pole position $\omega_P(\bp)$ being the solution of Eq.~\eqref{eq:propinv2},
while 
the dotted lines stand for the in-vacuum dispersion relation of the $\eta'$ meson, that is $\omega=\sqrt{m_{\eta'}^2+p^2}$.
One sees that the spectral functions have a peak along with the line of $\omega_R(\bp)$ and that the peak position $\omega_R(\bp)$ approaches the in-vacuum dispersion relation for larger $p$.
This means that the nuclear medium effect on the mass can be seen more significantly with smaller $p$.
Thus, in addition to the energy or invariant mass, the spatial momentum of the $\eta'$ meson $p$ characterizes the in-medium spectral function as well.

\section{Models}
\label{sec:models}
In this work, we evaluate the $\eta'$ self energy in the nuclear medium based on the $T\rho$ approximation~\eqref{eq:trho2} from the $\eta'N$ scattering amplitude. We utilize two models for the $T$-matrix to demonstrate possible scenarios of the nuclear modification of the $\eta'$ properties;
one is the \com{coupled channels model} and the other is the $N(1895)$-dominance model.
In the following subsections, we explain the details of two models for the $\eta'N$ scattering process.

\subsection{\com{Coupled channels model}}
\label{subsec:scattering}
\begin{figure}
 \centering
 \subfigure[]{\includegraphics[width=0.45\textwidth]{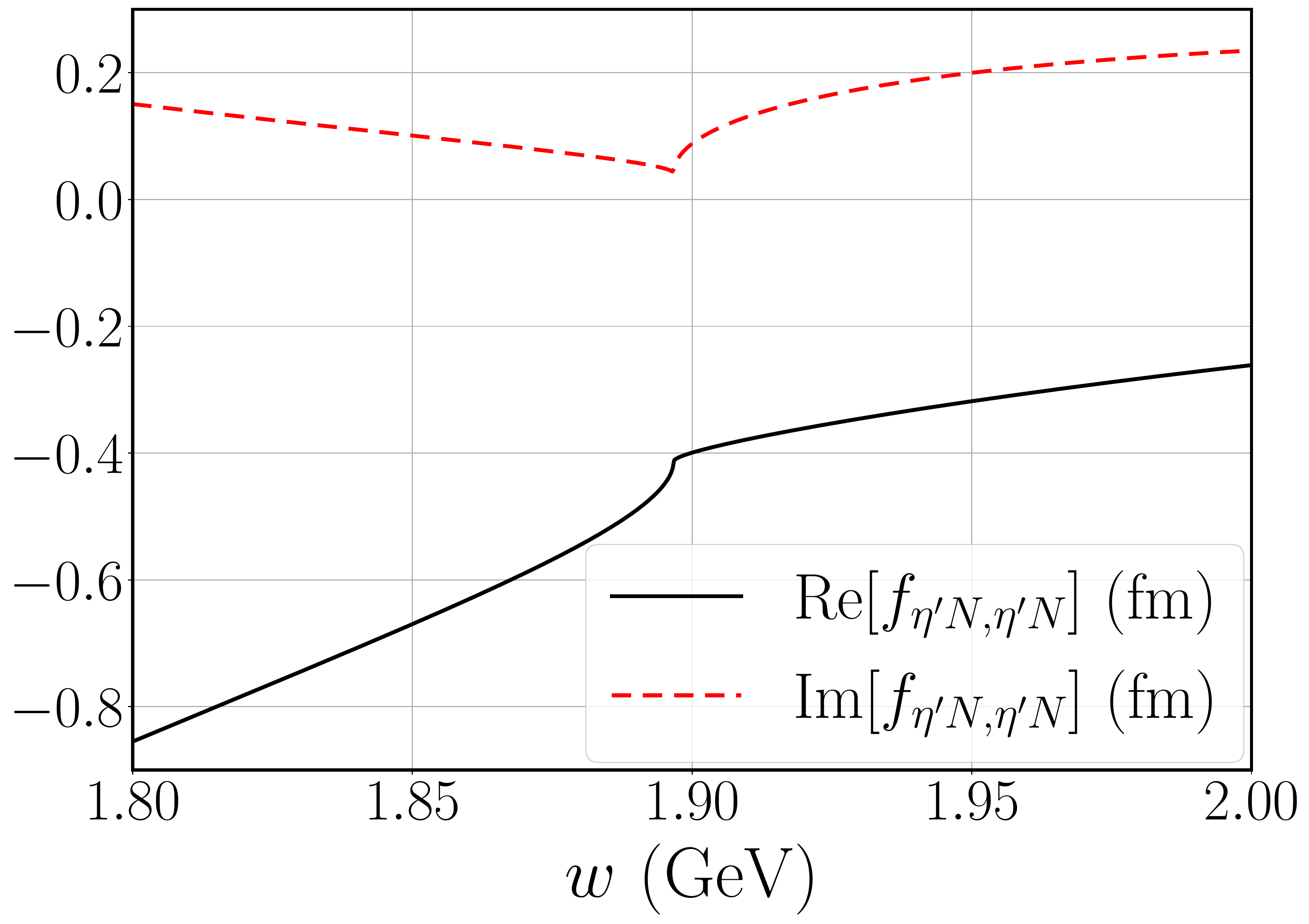}\label{fig:etapN_1}}
 \hspace{5mm}
 \subfigure[]{\includegraphics[width=0.45\textwidth]{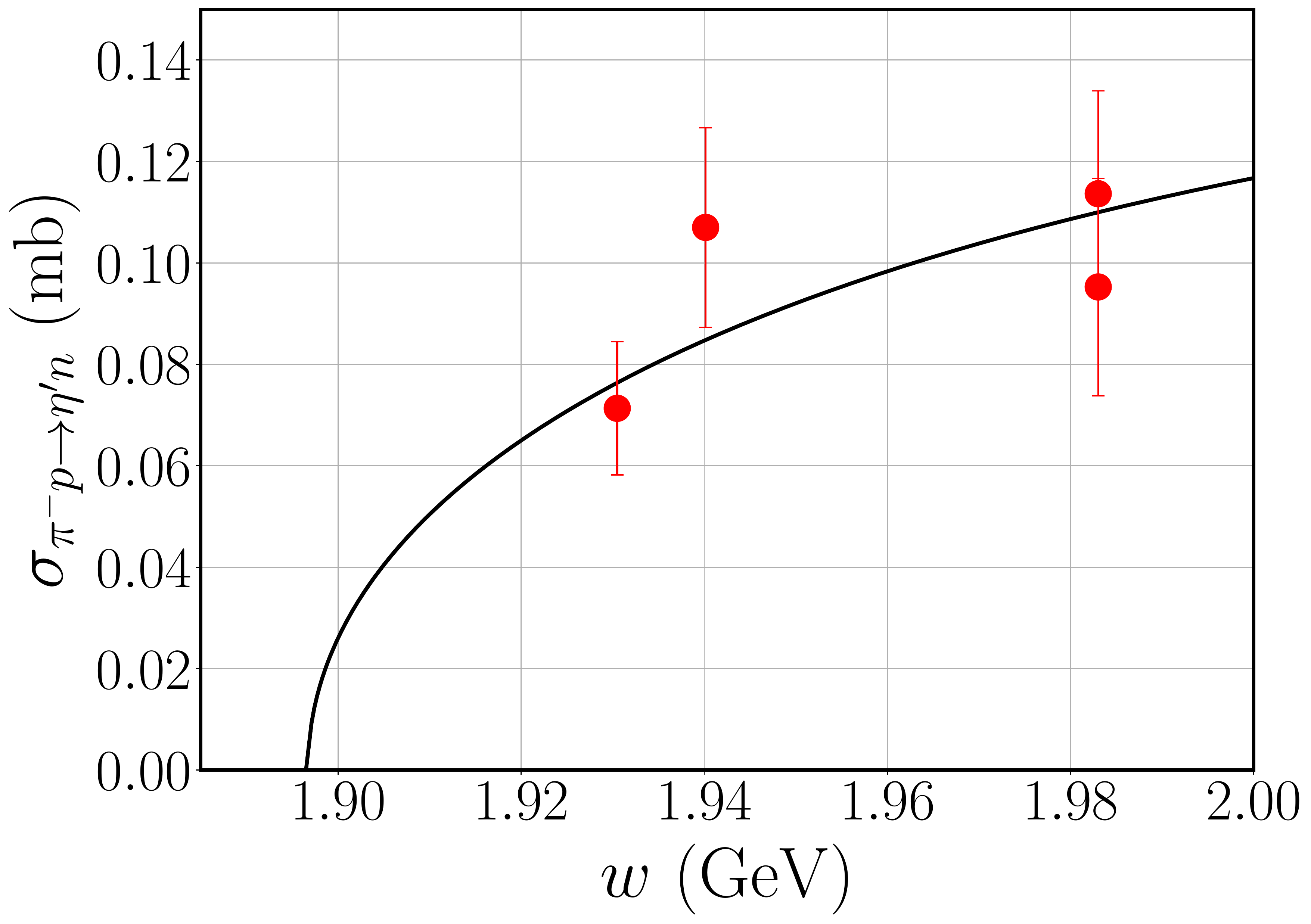}\label{fig:etapN_2}}
 \caption{(a)~Scattering amplitude of $\eta'N \to \eta'N$, $f_{\eta'N,\eta'N}$, calculated in the \com{coupled channels model} as a function of the c.m.~energy $w$ of the $\eta'N$ system.
 (b)~Total cross section of $\pi^-p\to\eta'n$ calculated in the \com{coupled channels model} as a function of the c.m.~energy $w$ of the $\eta'N$ system.
The points with error bar are the experimental data taken from Ref.~\cite{Schopper:1988vrx}.}
 \label{fig:etapN}
\end{figure}

Here let us explain the \com{coupled channels model}, which provides a more realistic $\eta'N$ scattering amplitude than the amplitude used in Sec.~\ref{sec:preliminaries}.
In this model, we utilize the $\eta'N$ two-body scattering amplitude developed in Ref.~\cite{Bruns:2019fwi} for the $\eta'N$ scattering $T$ matrix appearing in the in-medium $\eta'$ self energy~\eqref{eq:trho2}.
In Ref.~\cite{Bruns:2019fwi}, the scattering amplitude with $J^P=1/2^-$, $I=1/2$, $3/2$, and strangeness $S=0$ is studied from the view point of the meson-baryon dynamics. The meson-baryon scattering contains the $\eta'N$ channel together with coupled channels, $\pi N$, $\eta N$, $K\Lambda$ and $K\Sigma$ with $I=1/2$. 
The coupled-channel scattering amplitude of the $s$-wave meson-baryon pair, $f$, is represented as 
a complex $5\times 5$ matrix and is calculated by the Lippmann-Schwinger equation,
\begin{align}
 f=V+V\mG^{(0)} f,\label{eq:sleq}
\end{align}
where $\mG^{(0)}$ is the meson-baryon Green function with the free Hamiltonian and $V$ is 
the interaction kernel which governs the meson-baryon interaction. The interaction kernel is assumed to 
be a separable form given by
\begin{align}
 V_{mn}=g_{m}(p^2)v_{mn}(s)g_{n}(p'^2),
\end{align}
where the summation of the channel indices $n$ and $m$ is not taken, and $g_{n}(p^2)$ is the monopole form factor for channel $n$, $g_n(p^2)=1/(1+p^2/\alpha_n^2)$,  
{with the parameters $\alpha_{n}$ given in Ref.~\cite{Cieply:2013sya}.}
We use the parameter set for the model~A of Ref.~\cite{Bruns:2019fwi}.
The variable~$s$ is the Mandelstam variable and is given by the square of the two-body total energy in the c.m.~frame.
The $\eta'N$ scattering amplitude $f_{\eta'N,\eta'N}$ given by Eq.~\eqref{eq:sleq} and the $T$ matrix $T_{\eta'N}(\sqrt s)$ which is necessary for the evaluation of the in-medium $\eta'$ self energy~\eqref{eq:trho2} are related with 
\begin{align}
T_{\eta'N}(\sqrt s)=-\frac{8\pi \sqrt{s}}{E_N+m_N}f_{\eta'N,\eta'N}(\sqrt s).
\end{align}
In Eq.~\eqref{eq:sleq}, a series of the processes involving the infinite number of the meson-baryon scattering given by $V$ is taken into account like in the chiral unitary approach for $\Lambda(1405)$~\cite{Kaiser:1995eg,Oset:1997it,Oller:2000fj,Hyodo:2002pk,Hyodo:2003qa,Hyodo:2011ur}.
The interaction kernel $v$ is given by the chiral $U(3)$ Lagrangian with nine pseudoscalar mesons and octet baryons~\cite{Borasoy:2002mt} with the $s$-wave projection in the isospin basis.
See Ref.~\cite{Bruns:2019fwi} for the details.
Since the $s$-wave projected interaction kernel depends only on $s$ in the on-shell approximation, the Lippmann-Schwinger equation~\eqref{eq:sleq} can be solved in an algebraic way as
\begin{align}
 f=g(1-vG^{(0)})^{-1}vg,\label{eq:scat_eq}
\end{align}
where the diagonal element of the two-body loop function $G^{(0)}$  is given by
\begin{align}
\hspace{-15pt} G_n^{(0)}=\int\frac{d^3l}{(2\pi)^3}\frac{-4\pi g_n^2(l^2)}{k_n^2-\bl^2+i\epsilon}=\frac{(\alpha_n+ik_n)^2}{2\alpha_n}g_n^2(k_n^2),   \label{eq:gloopvacuum}
\end{align}
with the c.m.\ momentum for the $n$-th channel $k_n=\lambda^{1/2}(s,m_{n1}^2,m_{n2}^2)/(2\sqrt{s})$ for  
the masses of the meson and baryon, $m_{n1}$ and $m_{n2}$, respectively.
The calculated $\eta'N$ scattering amplitude $f_{\eta'N,\eta'N}$ and the $\pi^-p\to\eta'n$ cross section are shown in Fig.~\ref{fig:etapN}.
With this scattering amplitude, the $s$-wave $\eta'N$ scattering length $a_{\eta'N}$ is evaluated to be $a_{\eta'N}=(-0.41+i0.04)~\fm$ with the negative real part corresponding to the repulsive sign.

\subsection{$N(1895)$-dominance model}
\label{subsec:Nstdominance}
In this subsection, we explain the $N(1895)$-dominance model for the $\eta'N$ scattering amplitude to evaluate the in-medium $\eta'$ self energy. 
The $N(1895)$ resonance has $J^P=1/2^-$ and couples to $\eta'N$ in $s$ wave. The Review of Particle Physics~\cite{ParticleDataGroup:2022pth} reports $N(1895)$ to be almost at the $\eta'N$ threshold, and
an isobar model analysis EtaMAID2018~\cite{Tiator:2018heh} finds this resonance with the Breit-Wigner mass 
$m_{N^*}=1.8944~\gev$ and width $\Gamma_{N^*}=0.0707~\gev$. Since $N(1895)$ is located close to the $\eta'N$ threshold, this resonance may have an impact on the in-medium $\eta'$ properties. 
Actually, the $N(1880)$ and $N(1900)$ resonances also exist near the $\eta'N$ threshold. These resonances, however, have $J^P=1/2^+$ and $3/2^+$ and couple to the $\eta'N$ channel with the 
{$p$- and $f$-waves,}
respectively. Thus, they may give less dominant contribution compared with the $s$ wave resonance near the $\eta'N$ threshold. We do not consider the $N(1535)$ resonance, because the coupling of $N(1535)$ to the $\eta'N$ channel is not found in the global analysis of the $\eta N$ and $\eta' N$ amplitudes performed by EtaMAID2018~\cite{Tiator:2018heh}, although some contributions from $N(1535)$ to the $\eta'N$ amplitude may be expected as reported in Refs.~\cite{Cao:2008st,Zhong:2011ht}.
The study of the $\eta'N$ process in the meson-baryon scattering model developed in Ref.~\cite{Bruns:2019fwi} obtains  no $N^*$ resonance which can be associated with $N(1895)$. Nevertheless, Ref.~\cite{Bruns:2019fwi} mentions that the possibility to find $N(1895)$ as a $\eta'N$ dynamically generated state is not ruled out.

\begin{figure}
 \centering
 \subfigure[]{\includegraphics[width=3.7cm]{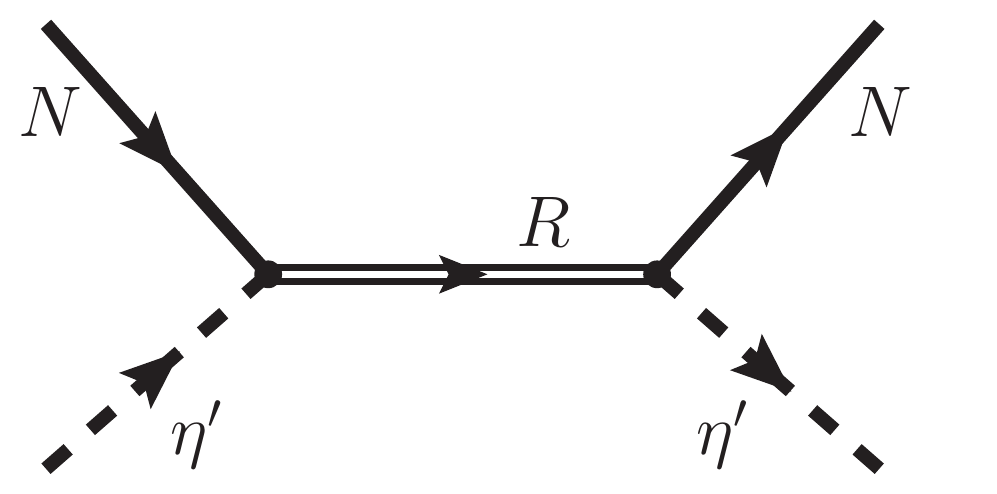}\label{fig:tree_res}}\qquad\quad
 \subfigure[]{\includegraphics[width=3.7cm]{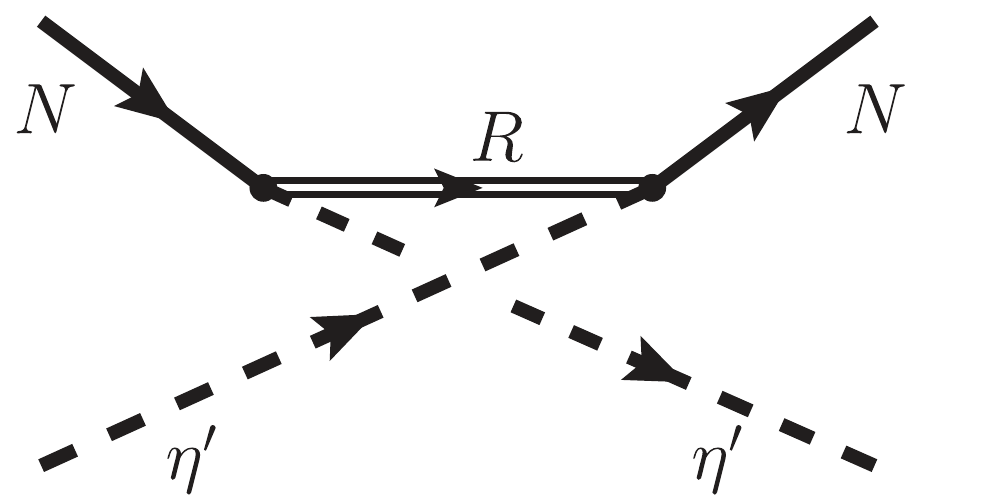}\label{fig:tree_res_cross}}
 \caption{Diagrams for the $N(1895)$ contribution 
 in the $s$-channel (a) and in the $u$-channel (b) of the $\eta'N$ scattering. 
 The double line represents the $N(1895)$ resonance.}
 \label{fig:Nsthole}
\end{figure}

Here, we investigate possible effect of the $N(1895)$ resonance on the in-medium $\eta'$ properties with a simple model,  basing the $s$-wave $\eta'N$ amplitude on the resonance dominance.
The $N^{*}$ dominance model for the in-medium $\eta$ self-energy was introduced in Ref.~\cite{Chiang:1990ft} and discussed in Refs.~\cite{Jido:2002yb,Nagahiro:2003iv,Nagahiro:2005gf,Jido:2008ng,Nagahiro:2008rj} where the $\eta$ self-energy is evaluated by the $T\rho$ approximation with the $\eta N$ scattering amplitude obtained by the $N(1535)$ resonance dominance.  
In the $N(1895)$-dominance model the $\eta'N$ scattering amplitude is described by the $N(1895)$ resonance as depicted in Fig.~\ref{fig:Nsthole}.
The diagram in Fig.~\ref{fig:tree_res} comes from the contribution of the resonance in the $s$ channel, while the crossed diagram shown in Fig.~\ref{fig:tree_res_cross} involves the resonance in the $u$ channel.
In the vicinity of the $\eta'N$ threshold, the $T$ matrix of the $\eta'N$ scattering in free space is written in the $\eta'N$ center-of-mass frame as
\begin{align}
  T_{\eta'N}(\sqrt{s})= & \frac{g_{\eta'N}^2}{\sqrt s-E_{N^*}+i\Gamma_{N^*}/2}
  \\ &
  +\frac{g_{\eta'N}^2}{-E_{\eta'}'+E_N-E_{N^*}+i\Gamma_{N^*}/2},\label{eq:tetapnres}
\end{align}
where the invariant mass is given by $\sqrt s = E_{\eta'}'+E_N$ with the $\eta'$ and nucleon energies, $E'_{\eta'}$ and $E_N$, respectively, $E_{N^*}$ and $\Gamma_{N^*}$ are the energy and width of the $N(1895)$ resonance, and $g_{\eta'N}$ is the coupling constant of $N(1895)$ to the $\eta'N$ channel. We use the energy-independent width for simplicity.
We obtain the in-medium $\eta'$ self energy $\Pi_{\eta'}$ by inserting the $T$ matrix~\eqref{eq:tetapnres} to Eq.~\eqref{eq:trho2}. There we take the rest frame of the nuclear medium with the $\eta'$ momentum $p_{\eta'}^\mu = (\omega, \bp)$ and the kinematical variables in Eq.~\eqref{eq:tetapnres} are given as
$E_{\eta'}'=\omega$, $E_{N} = m_{N}$, and $E_{N^{*}} = \sqrt{m_{N^{*}}^{2}+\bp^{2}}$ with 
the $N(1895)$ mass $m_{N^{*}}$. 
We use the values of the parameters given by the isobar-model analysis EtaMAID2018~\cite{Tiator:2018heh};
$m_{N^*}=1.8944~\gev$, $\Gamma_{N^*}=0.0707~\gev$, and $g_{\eta'N}= 1.4$.
The normalization of the coupling constant is adjusted so as to $(E_N(s)+m_N)/E_{N^*}= 1$ in the vicinity of the $\eta'N$ threshold.
With these resonance parameters, the $\eta'N$ scattering length obtained from Eq.~\eqref{eq:tetapnres} is found to be $(-0.02+i0.43)~\fm$. This value is close to the one extracted from the low-energy $pp\to pp\eta'$ process~\cite{Czerwinski:2014yot} and its real part is small.
This does not necessarily means, however, that the in-medium modification of the $\eta'$ spectral function 
{could be insignificant,}
because the $\eta'$ self-energy has strong energy dependence due to the resonance contribution to the $\eta'N$ amplitude.  
In the $N(1895)$ dominance model, 
we expect as sufficient medium effects on the $\eta'$ meson as that on the $\eta$ meson with the $N(1535)$ dominance, because the value of the coupling constant $g_{\eta'N}= 1.4$ is comparable with that of the $N(1535)$ resonance to the $\eta N$ channel, which is found to be about $2$ from the resonance partial decay width to the $\eta N$ channel~\cite{Nagahiro:2008rj}.
In the present work, we do not consider possible in-medium modifications of the $N(1895)$ quantities appearing in the amplitude, such as, the mass, width and coupling constant, for simplicity.

The mass parameter of $N(1895)$ used in this model, $m_{N^*}$, is given in Ref.~\cite{Tiator:2018heh}, which is slightly below the $\eta'N$ threshold. The $N(1895)$ mass can be above or below the $\eta'N$ threshold, if one takes the uncertainty 
{of the $N(1895)$ mass}
given in Review of Particle Physics~\cite{ParticleDataGroup:2022pth} seriously.
We will consider also the case of the $N(1895)$ mass above the $\eta'N$ threshold by changing the mass parameter $m_{N^*}$ to be $1.906~\gev$, which is obtained in the analysis of Ref.~\cite{Anisovich:2017bsk}, in order to see how the in-medium $\eta'$ spectral function changes compared with the one evaluated with $m_{N^*}<m_{\eta'}+m_N$. 

Before we move to the numerical results of 
{the spectral function of}
the in-medium $\eta'$ meson, we make a short remark on the poles of the in-medium $\eta'$ propagator in the $N(1895)$-dominance model.
With the in-medium $\eta'$ self energy obtained with Eq.~\eqref{eq:tetapnres},
the pole positions of the in-medium $\eta'$ propagator are obtained by the equation,
\begin{align}
 \left(\omega^2-\bp^2-m_{\eta'}^2\right)\left(\omega+m_N-E_{N^*}+\frac{i}{2}\Gamma_{N^*}\right)-g_{\eta'N}^2\rho=0,\label{eq:pole_resmodel}
\end{align}
where we have ignored the crossed-diagram contribution for simplicity.
With small $\rho$ for the $\eta'$ meson at rest, two solutions of Eq.~\eqref{eq:pole_resmodel} are found approximately as
\begin{align}
 \begin{split}
 \omega_P^{(1)}\sim&\, m_{\eta'}+\frac{g_{\eta'N}^2\rho/(2m_{\eta'})}{m_{\eta'}+m_N-m_{N^*}+\frac{i}{2}\Gamma_{N^*}},\\
 \omega_P^{(2)}\sim&\, m_{N^*}-m_N-\frac{i}{2}\Gamma_{N^*} \\ &
 -\frac{g_{\eta'N}^2\rho}{m_{\eta'}^2-(m_{N^*}-m_N-\frac{i}{2}\Gamma_{N^*})^2}.
 \end{split}  \label{eq:sol_approx}
\end{align}
The former pole corresponds to the $\eta'$ meson pole in vacuum, while the latter pole stems from the $N(1895)$ pole in the $\eta'N$ scattering amplitude. We call these poles, $\omega_P^{(1)}$ and $\omega_P^{(2)}$, $\eta'$ mode and $N(1895)$-hole mode, respectively. In vacuum the $N(1895)$-hole mode does not show up in the $\eta'$ spectral function, while at finite densities it appears in the $\eta'$ spectral function thanks to the coupling of the $N(1895)$ resonance to the $\eta'N$ channel.
We expect that these two poles provide two peaks in the in-medium $\eta'$ spectral function. 
The emergence of two modes due to the coupling of the meson to the nucleon excited state in the nuclear medium has been discussed, for example, in the study of the in-medium properties of the $\eta$ meson which couples to $N(1535)$ in $s$ wave~\cite{Jido:2008ng}.

\section{Results}
\label{sec:results}
In this section we show our result of the calculation for the in-medium $\eta'$ spectral function and the in-medium properties of the $\eta'$ meson. 
We consider two models for the $T$-matrix, the \com{coupled channels model} and the $N(1895)$-dominance model, as discussed in the previous section. 
In the \com{coupled channels model}, the scattering amplitude is constructed based on a meson-baryon coupled channels approach~\cite{Bruns:2019fwi} and the model parameters are determined phenomenologically. This model produces the total cross section of the $\eta'$ production in $\pi^{-} p \to \eta' n$. The $N(1895)$-model is considered as a more theoretical description of the $\eta'N$ scattering amplitude. 
{In this model, the $\eta'N$ scattering amplitude is described by the propagation of the $N(1895)$ nucleon resonance that is located just below the $\eta'N$ threshold and couples to the $\eta'N$ channel in $s$-wave.} 
With this model, we investigate possible signals if the nucleon resonance takes a significant role for the $\eta'$ meson in the nuclear medium. 
These two models provide different features of the spectral functions thanks to the different energy dependence of the $T$-matrix in these models. First we show our results for the $\eta'$ meson at rest in the nuclear matter, and then we consider the $\eta'$ meson with a finite spatial momentum in each model.

\subsection{\com{Coupled channels model}}
\label{subsusbsec:etapN}
Let us consider first 
the $\eta'$ meson at rest in the nuclear medium.
\begin{figure}
 \centering
 \subfigure[]{\includegraphics[width=0.43\textwidth]{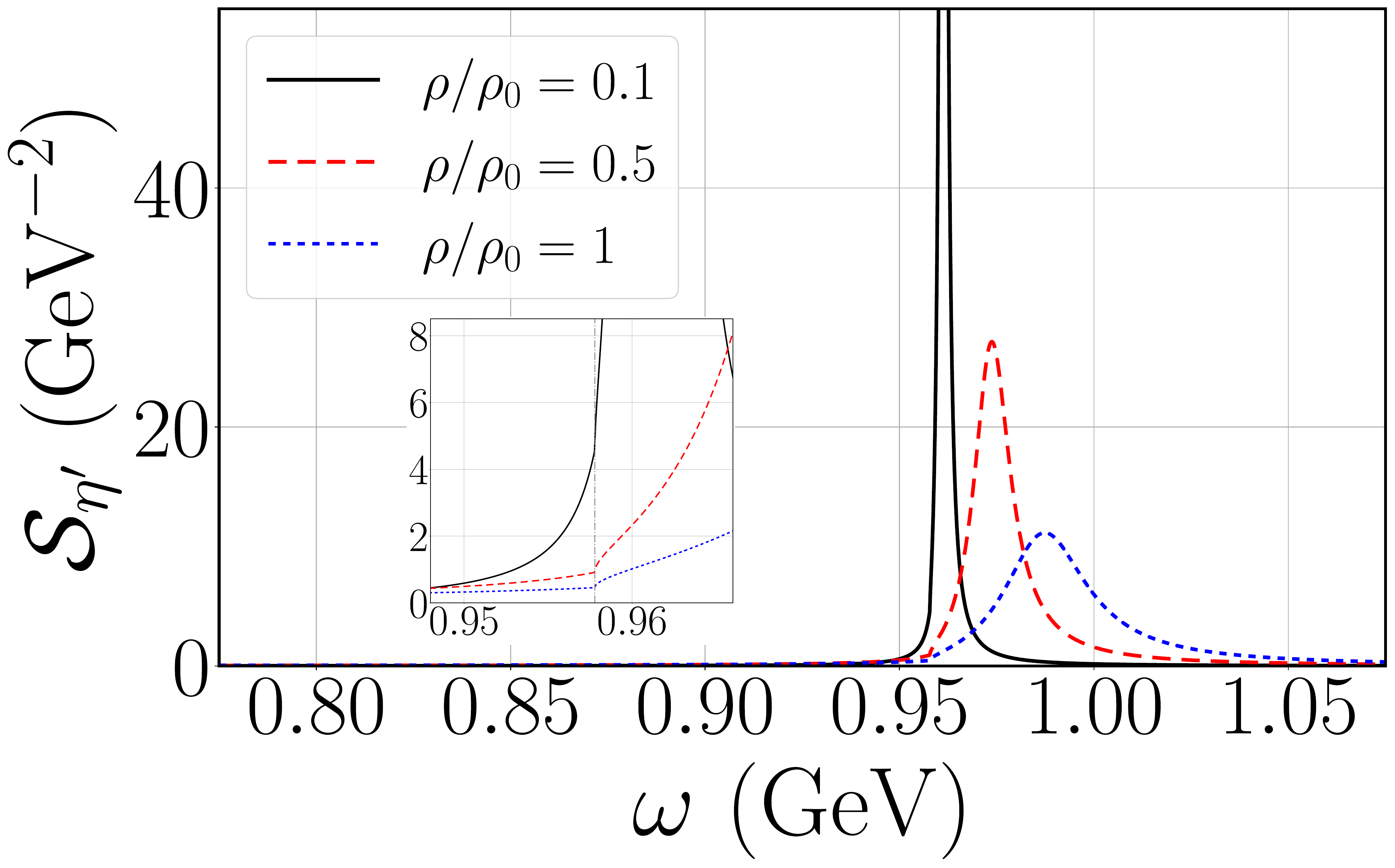}\label{fig:spec_11}}
 \subfigure[]{\includegraphics[width=0.43\textwidth]{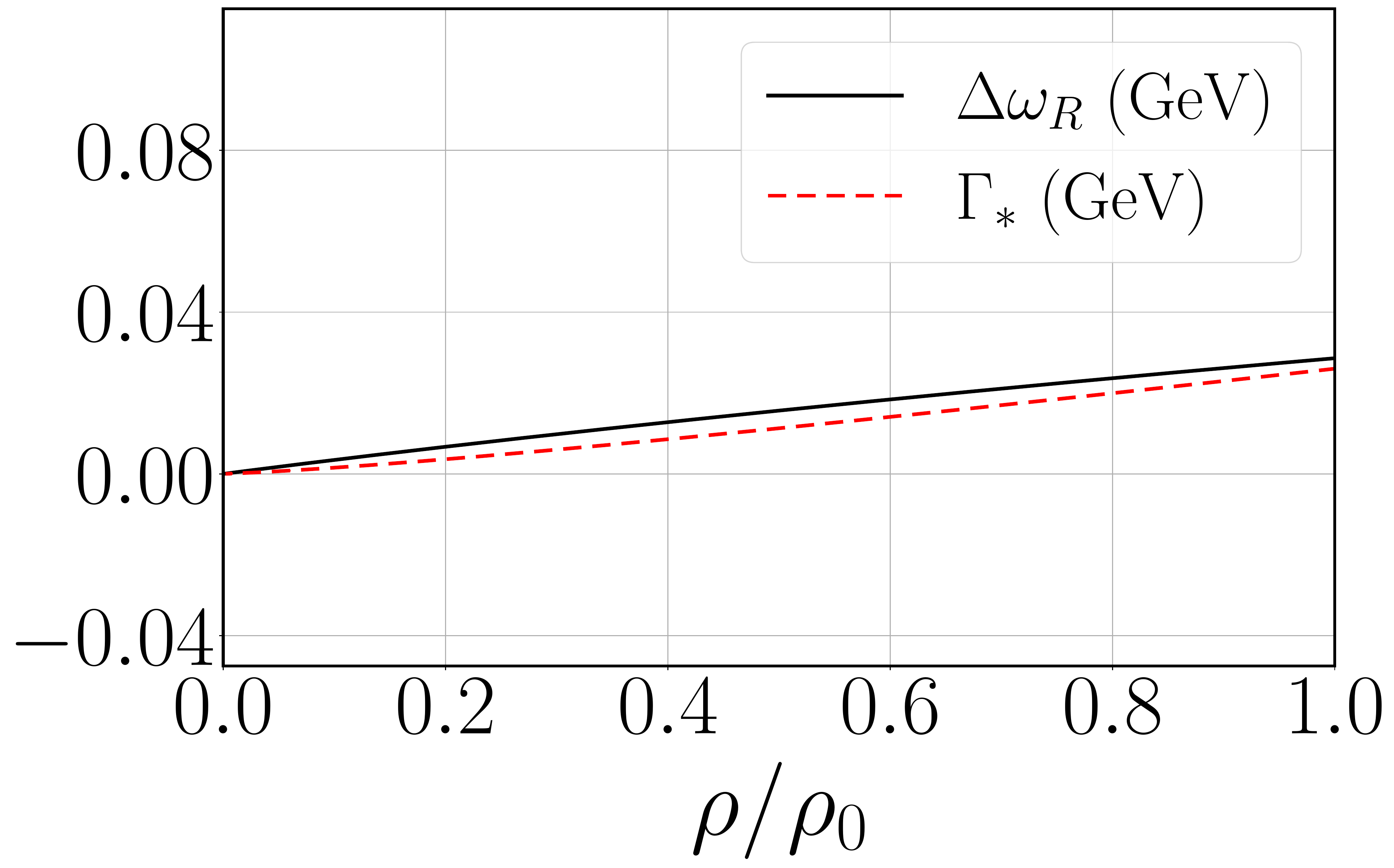}\label{fig:scat_mw_1}}
 \subfigure[]{\includegraphics[width=0.43\textwidth]{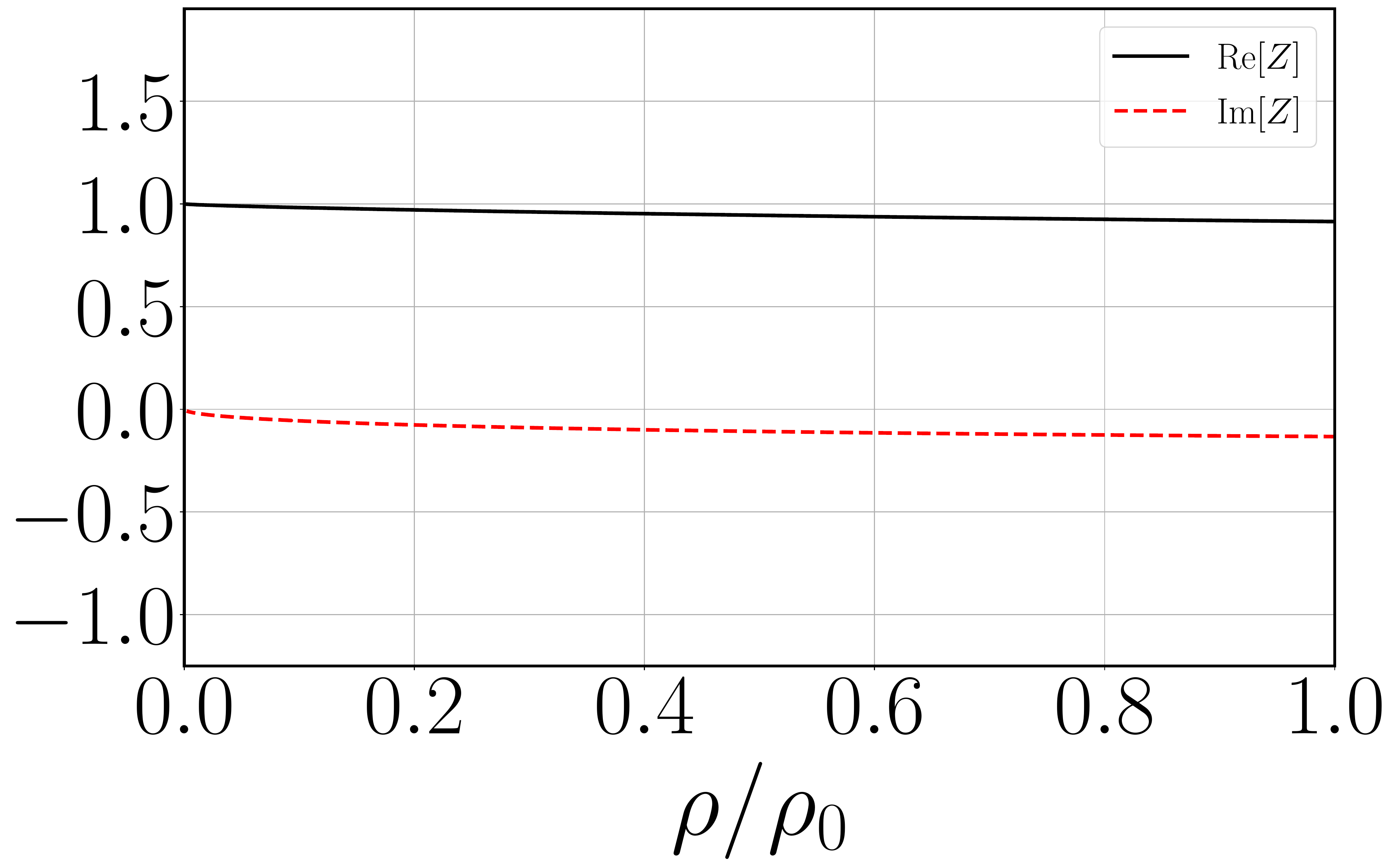}\label{fig:scat_res_1}}
 \caption{In-medium properties of the $\eta'$ meson at rest in the nuclear medium calculated with the \com{coupled channels model}:
(a)~the spectral functions $\mS_{\eta'}$ as functions of $\omega$ for nuclear densities $\rho=0.1\rho_{0},\, 0.5 \rho_{0},\, 1.0\rho_{0}$. The inserted figure is the enlarged view of the plots in the vicinity of $\omega=m_{\eta'}$. (b)~the in-medium mass modification $\Delta\omega_R=\omega_R-m_{\eta'}$ and in-medium $\eta'$ width $\Gamma_*$ in units of GeV as a function of density $\rho$. (c)~Real and imaginary parts of the wave function renormalization $Z$ as functions of density $\rho$.}
 \label{fig:spec_2}
\end{figure}
In Fig.~\ref{fig:spec_11}, we show the spectral function $\mS_{\eta'}$ as a function of the $\eta'$ energy $\omega$ with three fixed nuclear densities $\rho=0.1\rho_{0},\, 0.5 \rho_{0},\, 1.0\rho_{0}$ where $\rho_{0}$ is the normal nuclear density.
In this figure, we find that the peak position of the spectral function shifts towards higher energy from the in-vacuum $\eta'$ mass $m_{\eta'} =0.958~\gev$ for higher densities. It is also found that the width of the peaks gets wider for higher density and it reaches a few tens~MeV at $\rho = \rho_{0}$.
The high energy shift of the peak position is the consequence of 
the negative real part of the scattering length, $a_{\eta'N}=(-0.41+i0.04)~\fm$, as discussed in Sec.~\ref{sec:preliminaries}.
We make a small comment on the threshold structure of the in-medium $\eta'$ spectral function.
As seen in Fig.~\ref{fig:spec_11}, a threshold behavior appears at $\omega = m_{\eta'}$, that is the in-vacuum threshold. This threshold behavior originates from the intermediate $\eta'N$ state in the coupled channels calculation because we do not take into account of the medium effect on the intermediate states there. In this work we do not consider such medium effects on the $\eta'$ self-energy, which are beyond the $T\rho$ approximation. Because there is no dynamically generated states in the coupled channels, we expect that the medium effects on the intermediate states are not significant and the shape of the spectral function does not suffer from the in-vacuum threshold behavior. If these medium effects are important, one should perform a self-consistent calculation by including the medium effects to the intermediate states. 

In order to discuss the peak position of the spectral function more quantitatively, we show the in-medium mass and width of the $\eta'$ meson, $\omega_R$ and $\Gamma_*$, in Fig.~\ref{fig:scat_mw_1}. The in-medium mass and width are obtained from the pole position $\omega_P$ of the in-medium $\eta'$ propagator as Eq.~\eqref{eq:MassWidth}. In the figure we plot the mass modification defined by \com{$\Delta \omega_{R} = \omega_R - m_{\eta'}$}. The mass and the width of the $\eta'$ meson in the nuclear medium increase monotonically as the density increase. 
The mass shift at the normal density is evaluated to be 30~MeV in this model. The in-medium width stems from the nuclear absorption and, thus, increases as the density increases. The in-medium width at the normal density is found to be 30~MeV. 
In the $T\rho$ approximation, only the one nucleon absorption is taken into account. In this \com{coupled channels model}, the transition from $\eta'N$ to the $\pi N$, $\eta N$, $K \Lambda$ and $K\Sigma$ channels are responsible for the nuclear absorption of the $\eta'$ meson in the medium. 
The $\eta'N$ channel can also contribute to the absorption channel when the $\eta'$ energy is larger than the in-vacuum $\eta'$ mass, $\omega > m_{\eta'}$, because the in-vacuum hadron masses are used for the intermediate states.
We plot the in-medium wave function renormalization $Z$ in Fig.~\ref{fig:scat_res_1}. This figure shows about 10\% reduction of $Z$ at the saturation density. The size of the modification is moderate compared with the pion. Reference~\cite{Goda:2013npa} suggested that the wave function renormalization for pions is enhanced about 50\% at the saturation density and that the large in-medium modification of the wave function renormalization can lead to the change of the decay properties of the meson in the nuclear medium.
In this model for the $\eta'$ meson, we do not expect significant change of the $\eta'$ decay properties due to the modification of the meson normalization by the nuclear medium.

\begin{figure}
 \centering
 \subfigure[]{\includegraphics[width=0.45\textwidth]{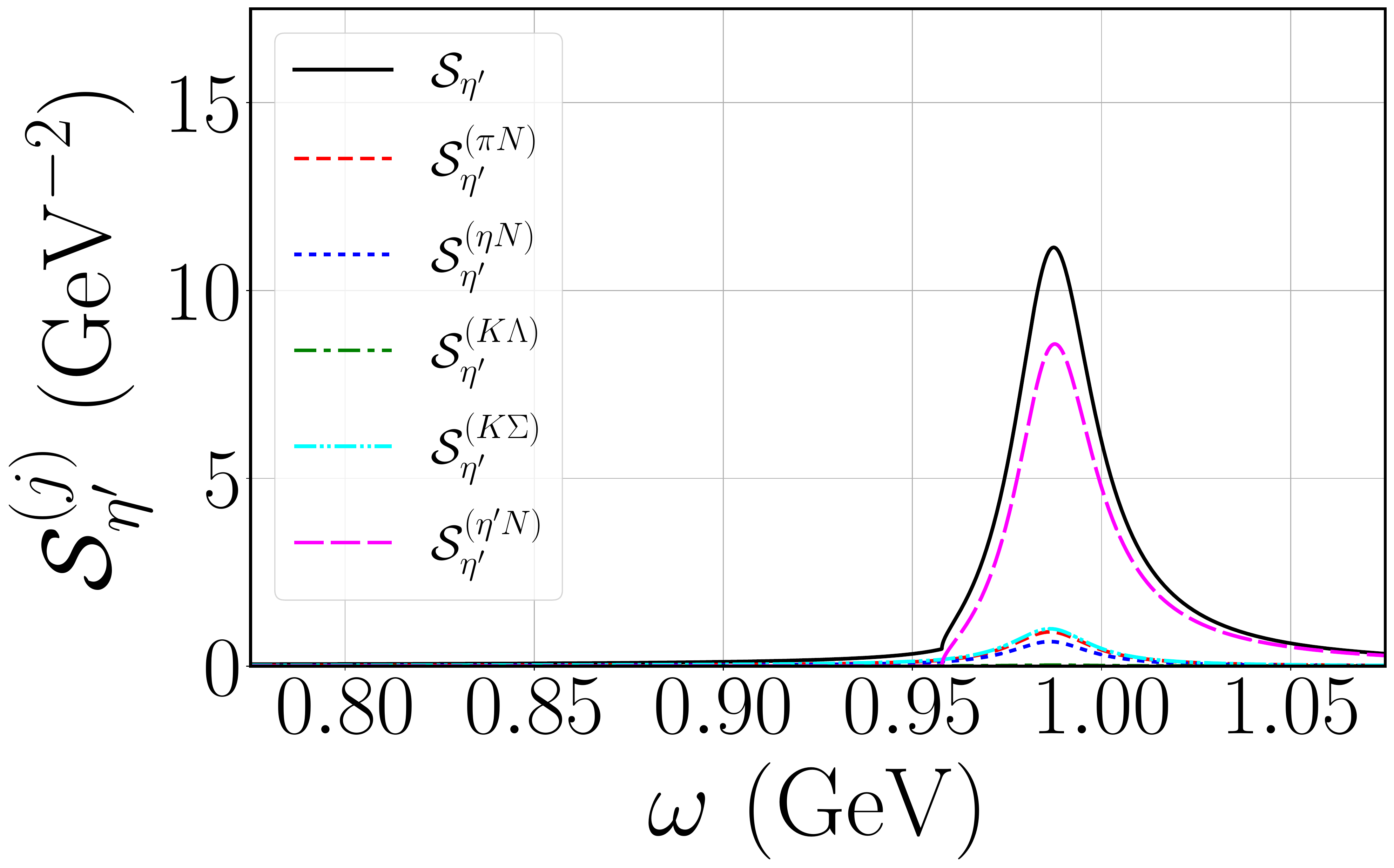}\label{fig:decomp1}}\qquad
 \subfigure[]{\includegraphics[width=0.45\textwidth]{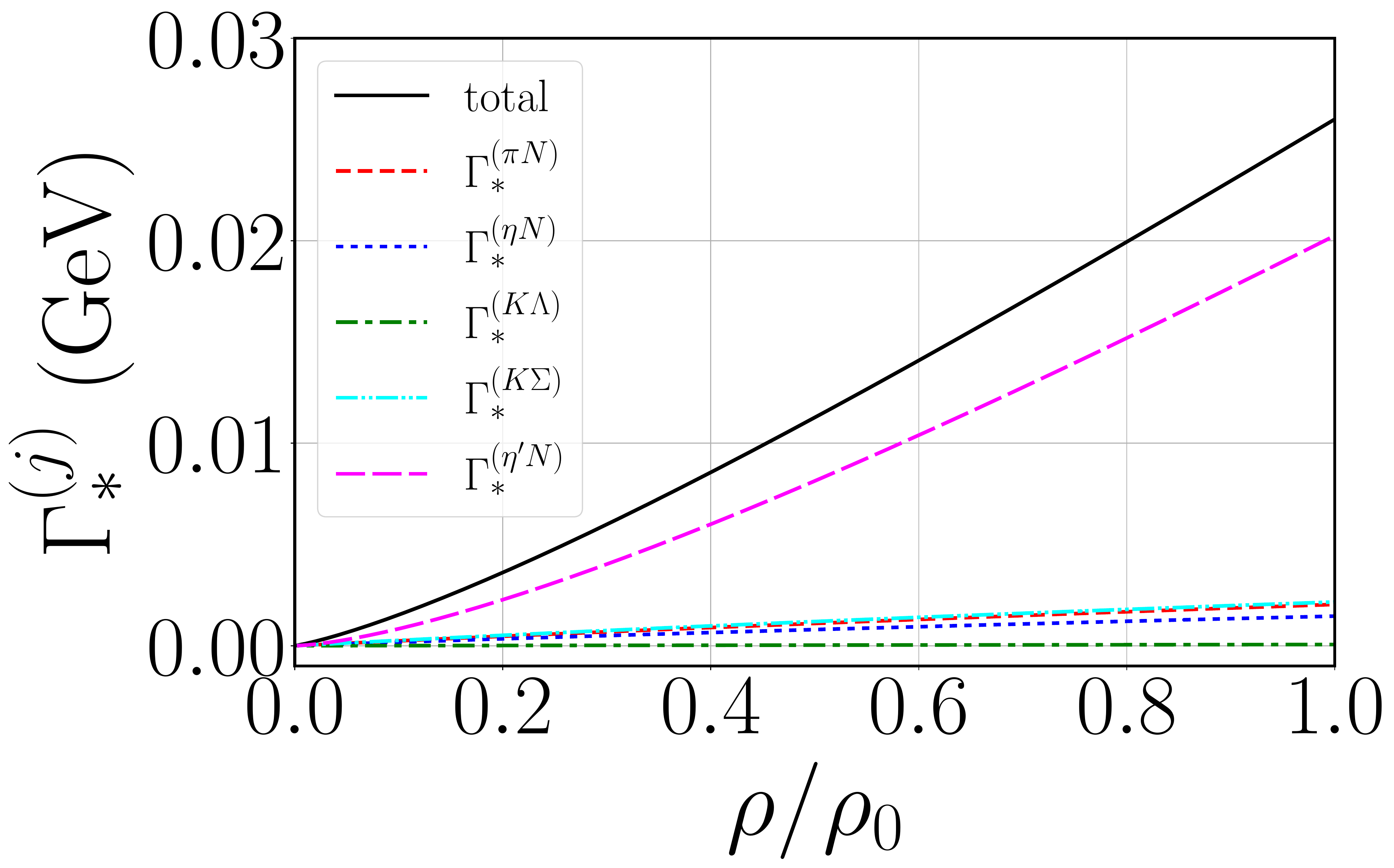}\label{fig:decomp2}}
 \caption{Decomposition of the spectral function and the in-medium width to each intermediate channel for the $\eta'$ meson at rest in the nuclear medium. 
 (a) Partial spectral functions defined in Eq.~\eqref{eq:partialSF} for $\rho = \rho_{0}$ as functions of the $\eta'$ energy. 
 (b) Partial widths defined in Eq.~\eqref{eq:partialwidth} as functions of $\rho/\rho_{0}$. }
 \label{fig:decomposition}
\end{figure}

The spectral function can be decomposed into the contribution from each intermediate state in the following way~\cite{Nagahiro:2008rj,Yamagata-Sekihara:2008qhf}; Using the free $\eta'$ propagator $D^{(0)}_{\eta'}(p) = (p^{2} - m_{\eta'}^{2} + i \epsilon)^{-1} $, the in-medium propagator is written as
$ D_{\eta'} =  D^{(0)} + D^{(0)} \Pi_{\eta'} D_{\eta'}$. 
It is known that the imaginary part of the Green function is decomposed to the two parts~\cite{Morimatsu:1985pf}:
\begin{align}
   {\rm Im}\left( D_{\eta'} \right)  = & (1 + D_{\eta}^{*} \Pi_{\eta'}^{*}) {\rm Im}\left(D^{(0)}_{\eta'} \right) (1+\Pi_{\eta'} D_{\eta'} ) \\ & + 
   D_{\eta'}^{*} {\rm Im}\left(\Pi_{\eta'}\right) D_{\eta'}  .
\end{align}
The first and second terms in the right hand side
are called escape part and conversion part, respectively. 
The escape part has the imaginary part of the free Green function and it provides the delta function for the in-vacuum dispersion relation. Because the in-medium $\eta'$ meson does not satisfy the in-vacuum dispersion relation, the escape part does not contribute to the spectral function for the infinite nuclear matter.  
For the conversion part, the imaginary part of the self-energy $\Pi_{\eta'}$ is written in terms of the imaginary part of the $\eta'N$ scattering amplitude in the $T\rho$ approximation~\eqref{eq:trho2}. 
Because the $\eta'N$ $T$-matrix is given by the coupled channels scattering, the conversion part can be decomposed into the contributions from each intermediate scattering state by using the optical theorem for the $\eta'N$ scattering amplitude: 
\begin{equation}
      {\rm Im}(T_{\eta'N}) =  \sum_{j} T_{\eta'N,j} \sigma_{j} T_{j,\eta'N}^{*},
\end{equation}
where $j$ is the channel index for the intermediate state, $T_{\eta'N,j}$ is the $T$-matrix for the transition of $\eta'N$ to the $j$ channel,  and $\sigma_{j}$ is the phase space factor. The phase space factor is given by
\begin{equation}
     \sigma_{j} = {\rm Im} \left( - \frac{E_{j}+m_{j}}{8\pi \sqrt s} \tilde G_{j}^{(0)} \right),
\end{equation}
with $\tilde G_{j}^{(0)} \equiv g_{j}^{-1}G_{j}^{(0)} g_{j}^{-1}$, where $G_{j}^{(0)}$ and $g_{j}$ are the free loop function~\eqref{eq:gloopvacuum} and the form factor for channel $j$, respectively.\footnote{The reason that we need the inverse of the form factor here is that the Lippmann Schwinger equation for $f$ is given by Eq.~\eqref{eq:sleq} and it reads $f = g v g+ g v g \tilde G^{(0)} f$. This implies that $\tilde G^{(0)}$ guarantees the unitarity of the scattering matrix.}
With these decomposition, we define the partial spectral function and partial in-medium width for channel $j$ as
\begin{align}
   S_{\eta'}^{(j)} = & - \frac{1}{\pi} D_{\eta'}^{*} \im \left( T_{\eta'N,j} \sigma_{j} T_{j,\eta'N}^{*} \right) \rho D_{\eta'} , \label{eq:partialSF}\\
   \Gamma_{*}^{(j)} = & - \frac{1}{\omega_{R}}  \left. \im \left( T_{\eta'N,j} \sigma_{j} T_{j,\eta'N}^{*} \right) \right|_{\omega=\omega_{P}} \rho, \label{eq:partialwidth}
\end{align}
respectively, where summation is {\it not} taken for the repeated index.  
We show the decompositions of the spectral function in Fig.~\ref{fig:decomp1} and the in-medium decay width in Fig.~\ref{fig:decomp2}. The spectral function is evaluated at the normal nuclear density $\rho = \rho_{0}$, while the partial decay widths are shown as functions of $\rho/\rho_{0}$.
In the figure, we find that the contribution of the $\eta'N$ channel dominates the spectral function and the in-medium width. 
In the \com{coupled channels model}, thanks to the repulsive nature of the scattering length the real part of the in-medium $\eta'$ mass is larger than the in-vacuum mass. Therefore, the $\eta'N$ channel is open at the pole position of the in-medium $\eta'$ propagator.  Channels other than $\eta'N$ give minor contributions. Among them the $\pi N$, $K\Sigma$ and $\eta N$ channels contribute, while the $K\Lambda$ channel is negligibly small. The fraction of the channel contributions is a consequence of the nature of the scattering amplitudes in the model applied here. The partial widths can be observed if one identifies the absorption channels of the $\eta'$ meson. 

In the \com{coupled channels model}, the modification of the $\eta'$ spectral function is relatively simple;
the nuclear medium effect emerges from the scattering of the $\eta'$ meson with the nucleons in the medium, and it causes the shift and the broadening of the peak in the in-medium $\eta'$ spectral function.
The wave function renormalization does not differ from unity so much.

\begin{figure}
 \centering
 \subfigure[]{\includegraphics[width=0.45\textwidth]{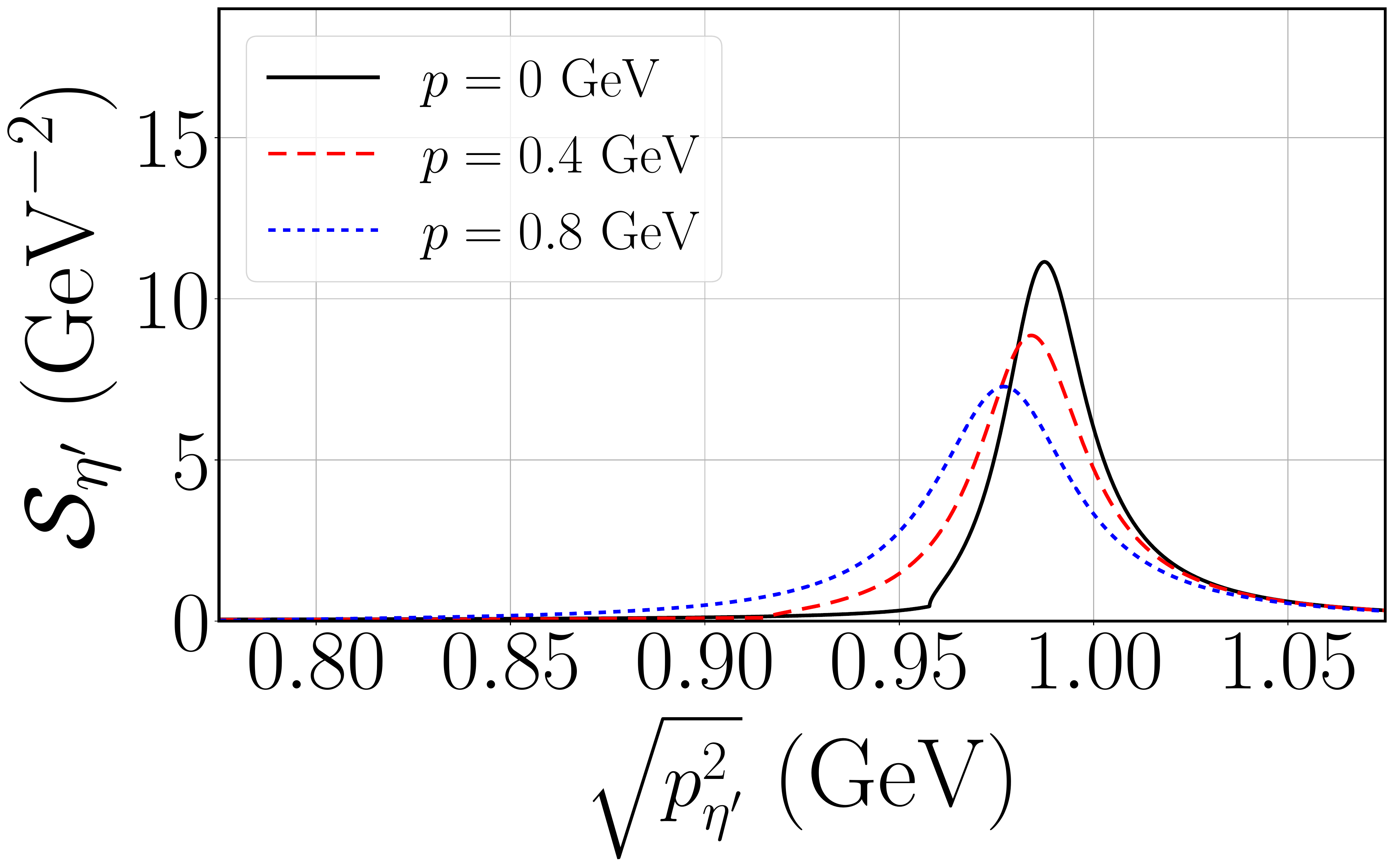}\label{fig:spec_momdep_1a}}
 \hspace{5mm}
 \subfigure[]{\includegraphics[width=0.45\textwidth]{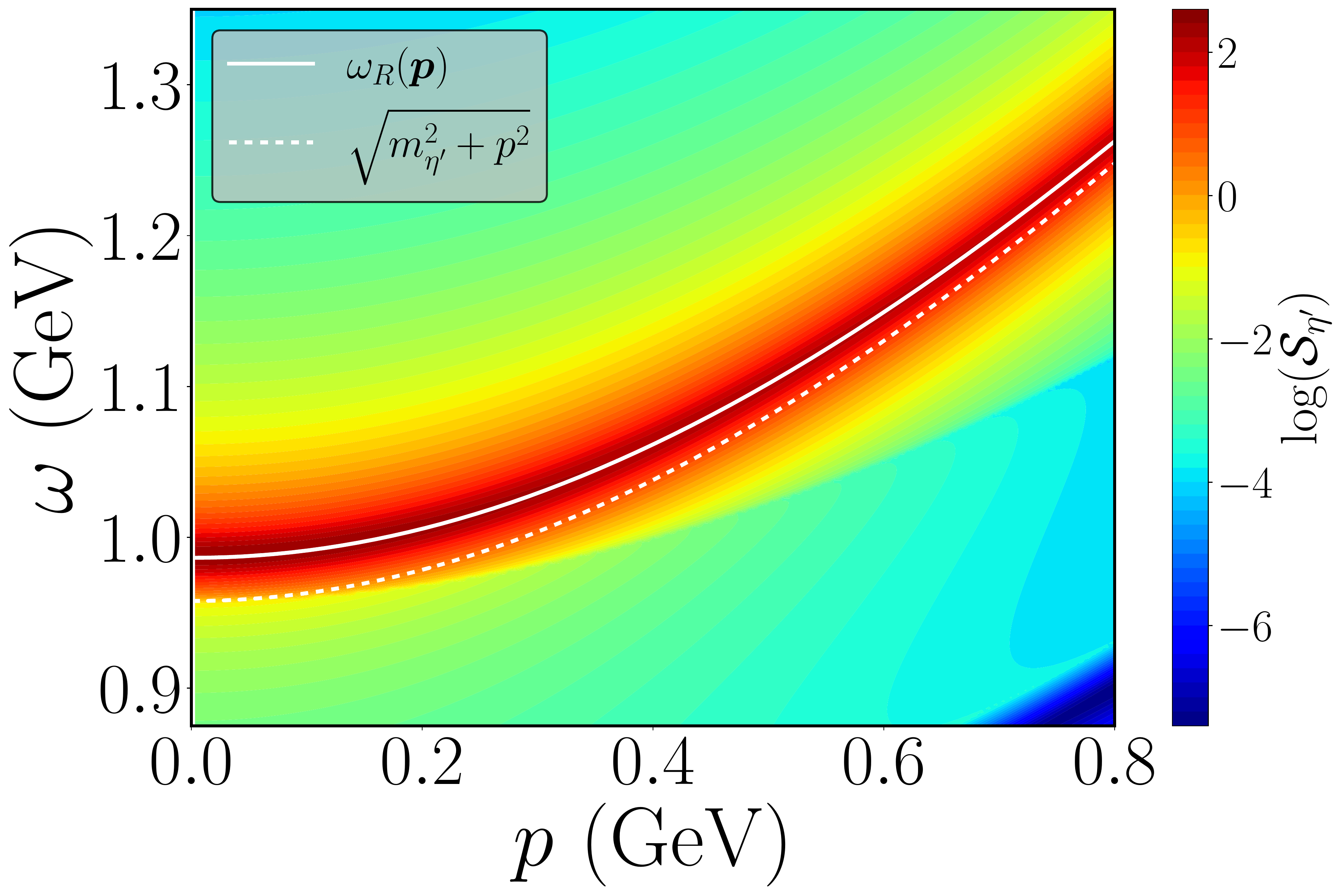}\label{fig:plot2da}}
 \caption{Momentum dependence of the spectral function at $\rho=\rho_0$ in the \com{coupled channels model}.
 (a)~The spectral functions with fixed $\eta'$ momenta $p = 0,\ 0.4,\ 0.8\ \gev$ are plotted as functions of the $\eta'$ invariant mass $\sqrt{p_{\eta'}^2}=\sqrt{\omega^2-\bp^2}$.
 (b)~A contour plot of the logarithm of the $\eta'$ spectral function in the $p$-$\omega$ plane at $\rho=\rho_{0}$ is shown. The solid and dotted lines are the in-medium dispersion relation $\omega_R(\bm{p})$ and the in-vacuum dispersion $\omega = \sqrt{m_{\eta'}^{2}+ \bp^{2}}$.
}
 \label{fig:spec_momdep_1}
\end{figure}

Finally, let us discuss the $\eta'$ momentum dependences of the spectral function for the $\eta'$ meson with
$p^{\mu}_{\eta'} = (\omega, \bp)$.
In Fig.~\ref{fig:spec_momdep_1} 
we show the spectral function as a function of the $\eta'$ invariant mass $\sqrt{p_{\eta'}^2}=\sqrt{\omega^2-\bp^2}$ at $\rho=\rho_0$ with fixed spatial
momenta $p = 0,\ 0.4,\ 0.8\ \gev$
and a contour plot of the logarithm of the $\eta'$ spectral function $\log\left(\mS_{\eta'}\right)$ at $\rho=\rho_{0}$ in the $p$-$\omega$ plane.
As seen in Fig.~\ref{fig:spec_momdep_1a}, the peak of the spectral function moves to lower energy for larger $\eta'$ momentum $p$. This is the same tendency as seen in
Fig.~\ref{fig:spec_mom_dem_3} for the effective-range approximation with the scattering length $\re(a_{\eta'N})<0$. There we have found that the peak position in the spectral function approaches the $\eta'$ mass in vacuum for finite $\eta'$ momenta. This is seen in the contour plot
shown in Fig.~\ref{fig:plot2da} 
as the two lines approach each other for larger $p$. For the finite $\eta'$ momenta, although the in-medium modification of the $\eta'$ mass gets less evident in the \com{coupled channels model}, the shift of the peak position for $p=0.8\ \gev$ is still about 10~$\mev$. This is not so small compared with the shift of the peak position for the $\eta'$ meson at rest.

\subsection{$N(1895)$-dominance model}
\label{subsusbsec:Nst}
\begin{figure}[h]
 \centering
 \subfigure[]{\includegraphics[width=0.45\textwidth]{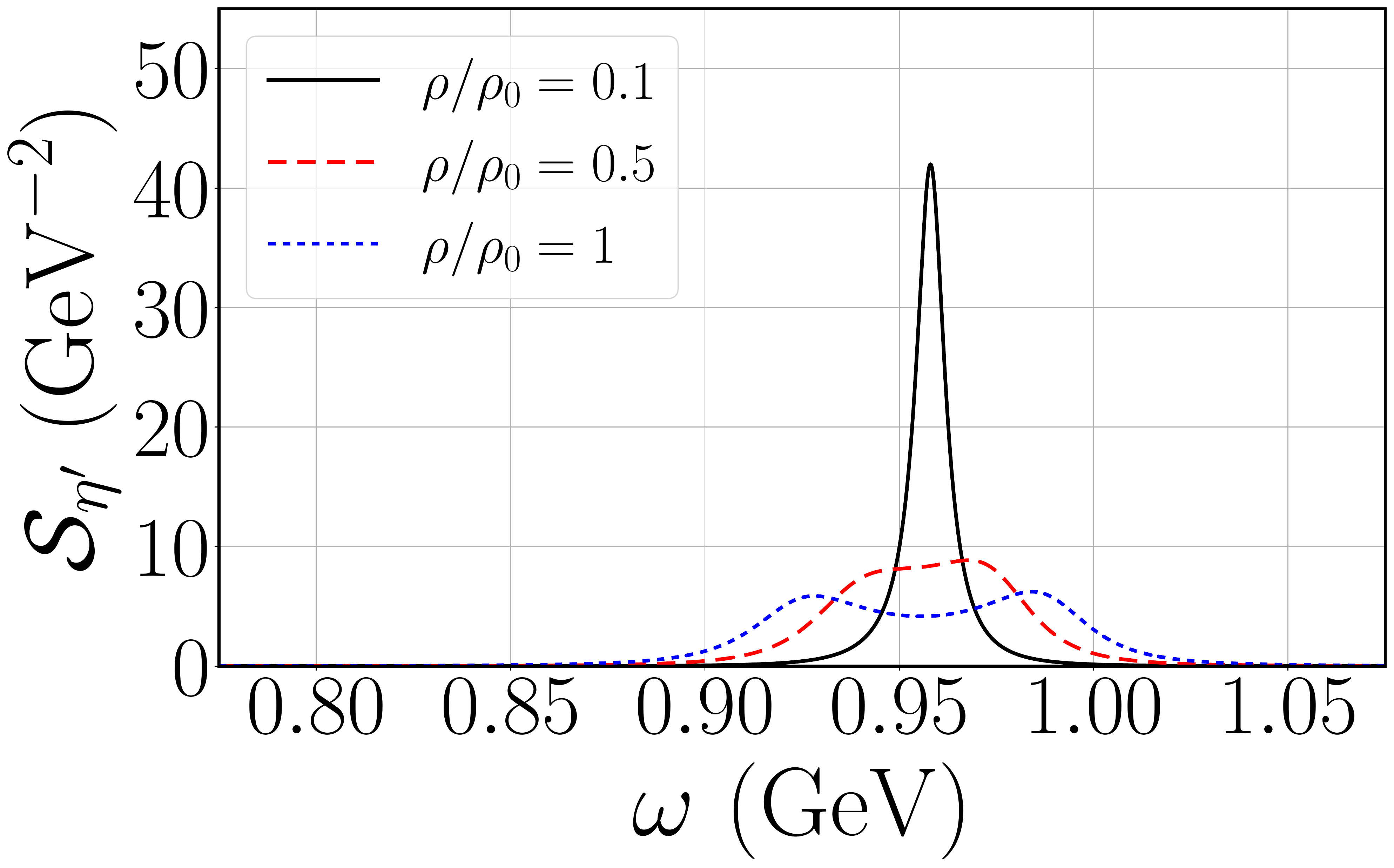}\label{fig:spec_22}}
 \subfigure[]{\includegraphics[width=0.45\textwidth]{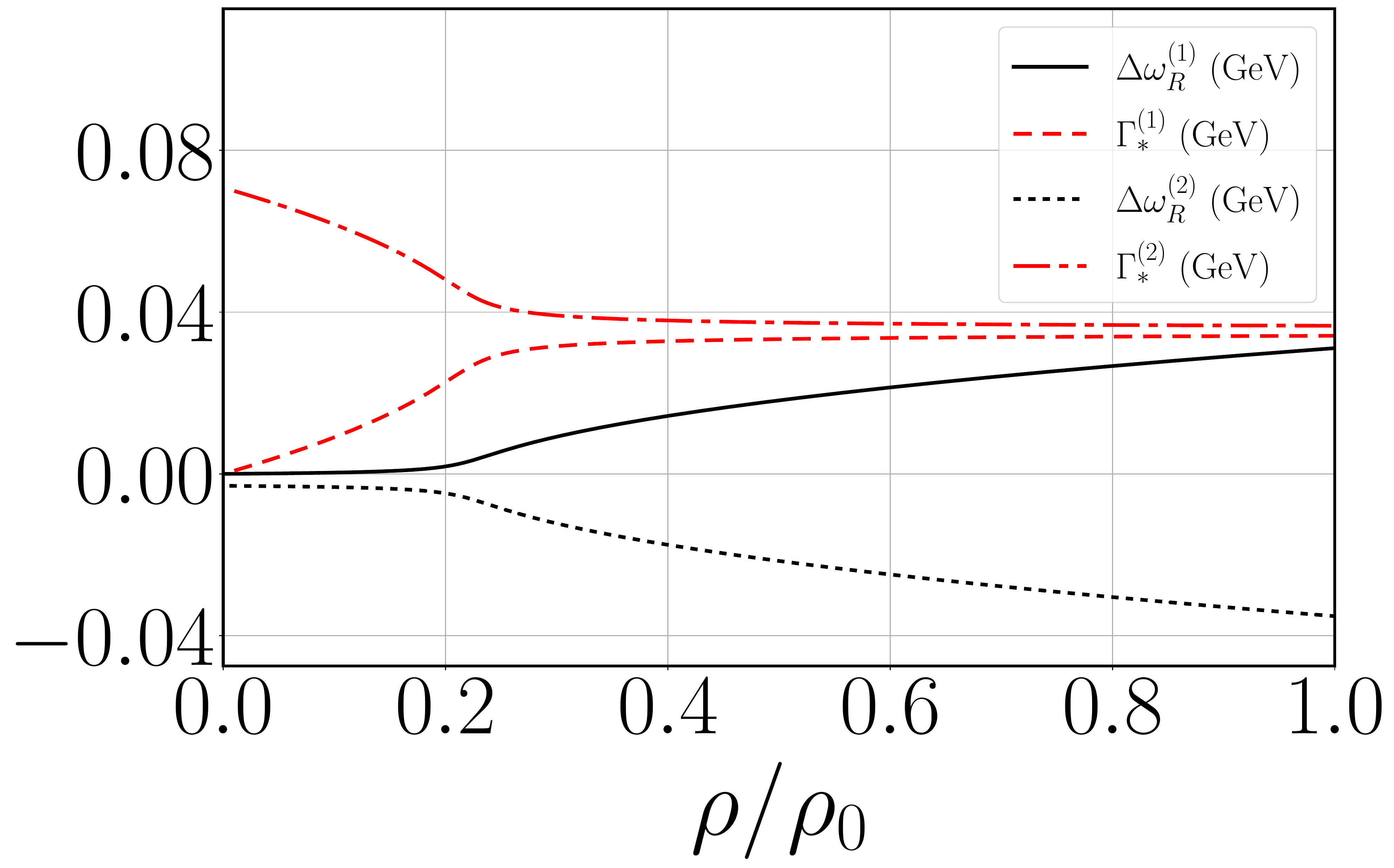}\label{fig:res_mw_res_1}}
 \subfigure[]{\includegraphics[width=0.45\textwidth]{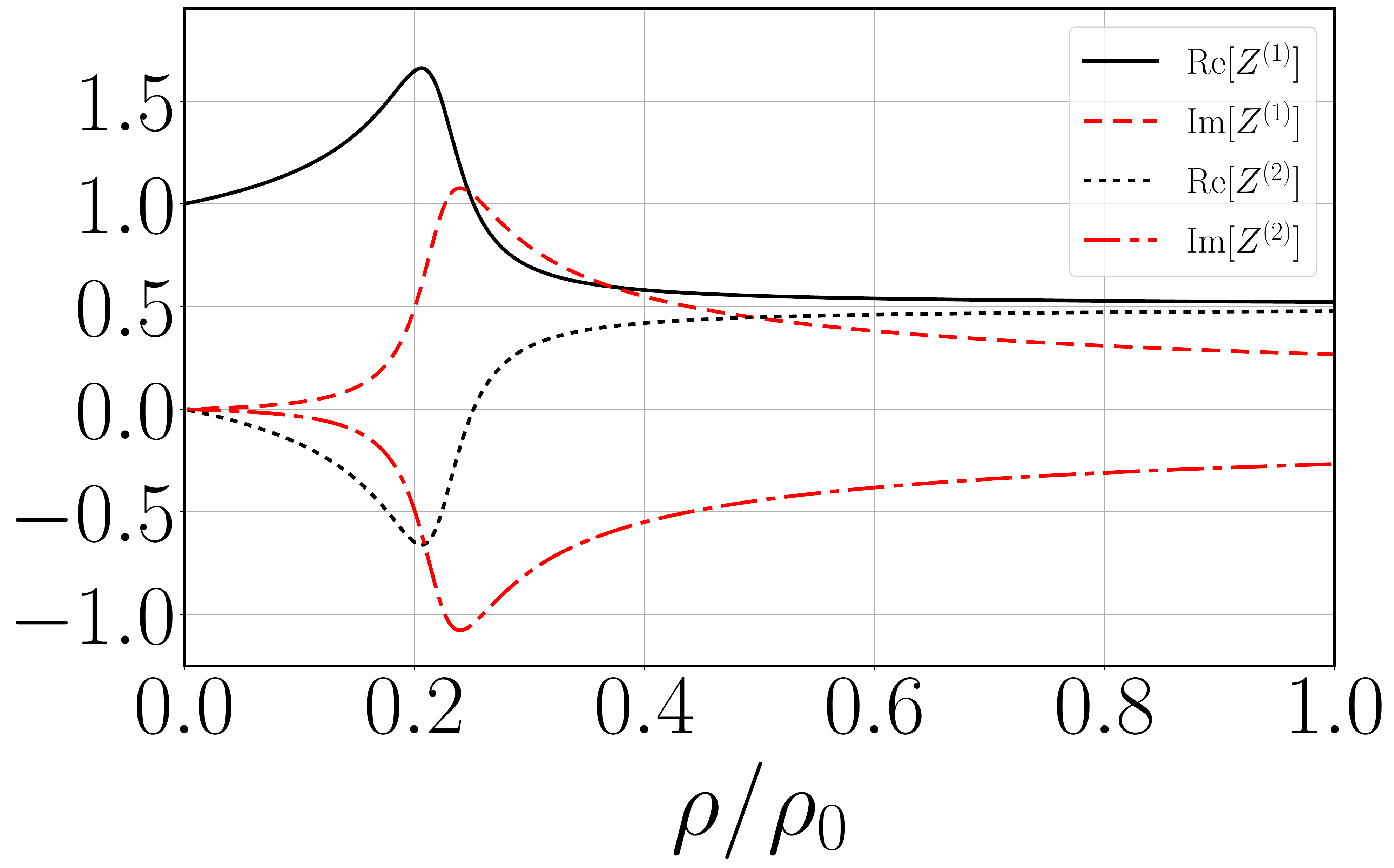}\label{fig:res_mw_res_2}}
 \caption{In-medium properties of the $\eta'$ meson at rest in the nuclear medium calculated with the $N(1895)$ resonance model:
(a)~the spectral functions $\mS_{\eta'}$ as functions of $\omega$ for nuclear densities $\rho=0.1\rho_{0},\, 0.5 \rho_{0},\, 1.0\rho_{0}$. (b)~the in-medium mass modifications $\Delta\omega_R=\omega_R-m_{\eta'}$ and in-medium $\eta'$ widths $\Gamma_*$ in units of GeV as functions of density $\rho$. (c)~Real and imaginary parts of the wave function renormalization $Z$ as functions of density $\rho$.
} \label{fig:res_mw_res}
\end{figure}
Now, we move to the results of the $N(1895)$-dominance model explained in Sec.~\ref{subsec:Nstdominance}.
In Fig.~\ref{fig:res_mw_res}, we present the spectral function, the in-medium mass and width, and the wave function renormalization for the $\eta'$ meson at rest in the nuclear medium. 
In Fig.~\ref{fig:spec_22}, we find that a single peak is located at the in-vacuum $\eta'$ mass for lower density $\rho=0.1 \rho_{0}$ and it splits into two peaks when the density increases. At the normal nuclear density $\rho=\rho_0$, the spectral function possesses two peaks located at $\pm 30~\mev$ away from the in-vacuum $\eta'$ mass with the width of a few tens $\mev$.
The positions of these peaks look almost symmetric against $\omega=m_{\eta'}$. 
Solving Eq.~\eqref{eq:propinv} for the $N(1895)$-dominance model to look for the poles of the in-medium $\eta'$ propagator, we obtain two solutions which correspond to 
two peaks of the spectral function. We name the pole having a larger (smaller) real part pole~$1$ (pole~$2$).
In Fig.~\ref{fig:res_mw_res_1} we show the density dependence of the pole positions by plotting the in-medium mass modifications $\Delta \omega_R^{(i)} = \omega_R^{(i)} - m_{\eta'}$ and the in-medium widths $\Gamma_*^{(i)}$ for two poles $i=1,2$. We find that their behavior changes around $\rho=0.2\rho_0$; below $\rho=0.2\rho_0$, the mass modifications $\Delta\omega_R^{(i)}$ are almost constant as they are in vacuum, while the in-medium widths $\Gamma_*^{(i)}$ rapidly change their magnitude in a few tens MeV. For $\rho > 0.2 \rho_{0}$, $\Delta\omega_R^{(1)}$ and $\Delta\omega_R^{(2)}$ split into higher and lower energies and their values in magnitude get increasing as the density increases, while the widths get almost constant with about 30 MeV. 

The density dependence of the wave function renormalizations $Z^{(i)}$ for $i=1,2$ are shown in Fig.~\ref{fig:res_mw_res_2}.
First of all, one finds that the wave function renormalization for pole~2, $Z^{(2)}$, is almost zero for $\rho < 0.2 \rho_{0}$. This implies that pole~2 little contribute to the spectral function. 
At the end of Sec.~\ref{subsec:Nstdominance}, we have discussed the appearance of two poles in the in-medium $\eta'$ propagator and their origins; one comes from the $\eta'$ mode and the other from the $N(1895)$-hole mode. 
Because pole 1 starts from the in-vacuum $\eta'$ mass and pole 2 begins at $m_{N*}-m_{N}$,
pole 1 and pole 2 contain dominantly the $\eta'$ mode and the $N(1895)$-hole mode, respectively, at low density. Thus, the clear peak in the spectral function at $\rho=0.1\rho_0$ in Fig.~\ref{fig:spec_22} is mainly attributed to the $\eta'$ mode. One sees that the behavior of the pole positions for lower density shown in Fig.~\ref{fig:res_mw_res_1} agrees with what we expect from the approximate pole positions~\eqref{eq:sol_approx}; the real part of $\omega_P^{(1)}$ moves to larger energies, while 
the real part of $\omega_P^{(2)}$ decreases as the density increases. The imaginary part of $\omega_P^{(2)}$ comes from the $N(1895)$ width.
Secondly one notices the characteristic peak structure around $\rho = 0.2 \rho_{0}$. The wave function renormalizations of both poles exhibit the peak structure at \com{$\rho=0.2\rho_0$} and their real parts approaches $0.5$ when the density is increased. This implies that strong cooperation of two modes to the spectral function takes place around
$\rho = 0.2 \rho_{0}$. Third,
for $\rho > 0.4 \rho_0$, $\re[Z^{(1)}]$ and $\re[Z^{(2)}]$ approach $0.5$. This indicates that the $\eta'$ and $N(1895)$-hole modes are largely mixed in these densities. The heights of two peaks in the spectral function at $\rho=\rho_0$ in Fig.~\ref{fig:spec_22} are close to each other since the size of $Z^{(1)}$ and $Z^{(2)}$ are similar at this density.
Although the density dependence of $Z^{(i)}$ looks not so simple, the sum of two residues is almost unity independently of the nuclear density, $Z^{(1)}+Z^{(2)}\sim 1$.
See, e.g., Refs.~\cite{Jido:2008ng,Nagahiro:2013hba} for the detailed discussion on the density dependence of the pole motion and its residue.

In this way, the in-medium $\eta'$ spectral function obtained with the $N(1895)$-dominance model shows different behavior from the one obtained with the \com{coupled channels model}. In the $N(1895)$-dominance model one finds two peaks coming from the $\eta'$ and $N(1895)$-hole modes for higher density $\rho$.
The shift of the pole positions and the widths of the peaks are a very similar size to those obtained with the \com{coupled channels model} at $\rho=\rho_0$, which is a few tens of $\mev$.

\begin{figure}
 \centering
 \includegraphics[width=0.45\textwidth]{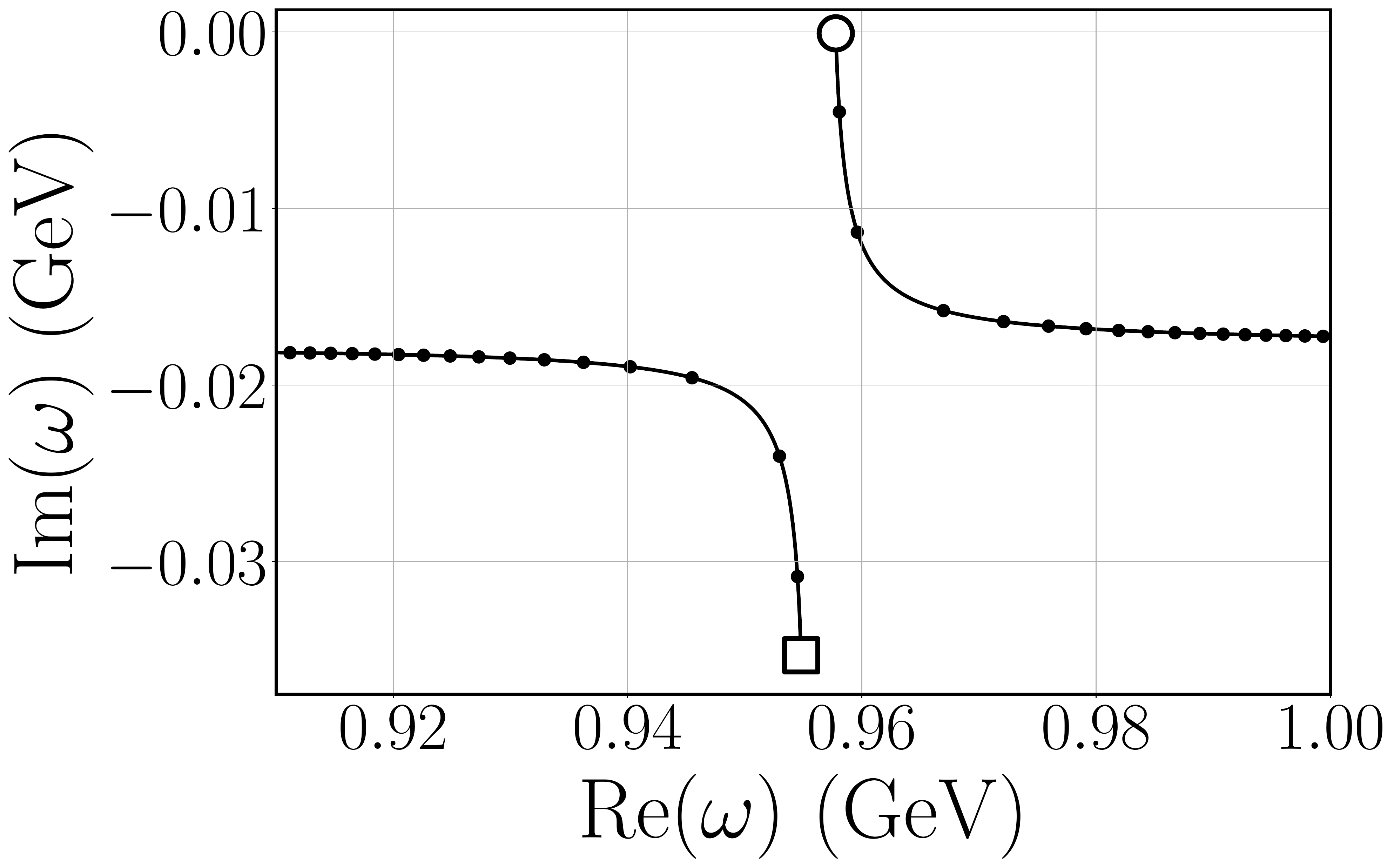}
 \caption{Pole trajectories $(\omega_R, - \Gamma_*/2)$ of the in-medium $\eta'$ propagator in the $N(1895)$-dominance model as the nuclear density $\rho$ varies. 
The open circle and square are the points given by $\omega=(m_{\eta'},0)$ and $\omega = (m_{N^*}-m_N,-\Gamma_{N^*}/2)$ which pole 1 and 2 approach in the limit of $\rho \to 0$,
respectively, and the dots are plotted at every $0.1 \rho_{0}$.
}
 \label{fig:poletrajectory1}
\end{figure}

\begin{figure}[t]
 \centering
 \includegraphics[width=0.45\textwidth]{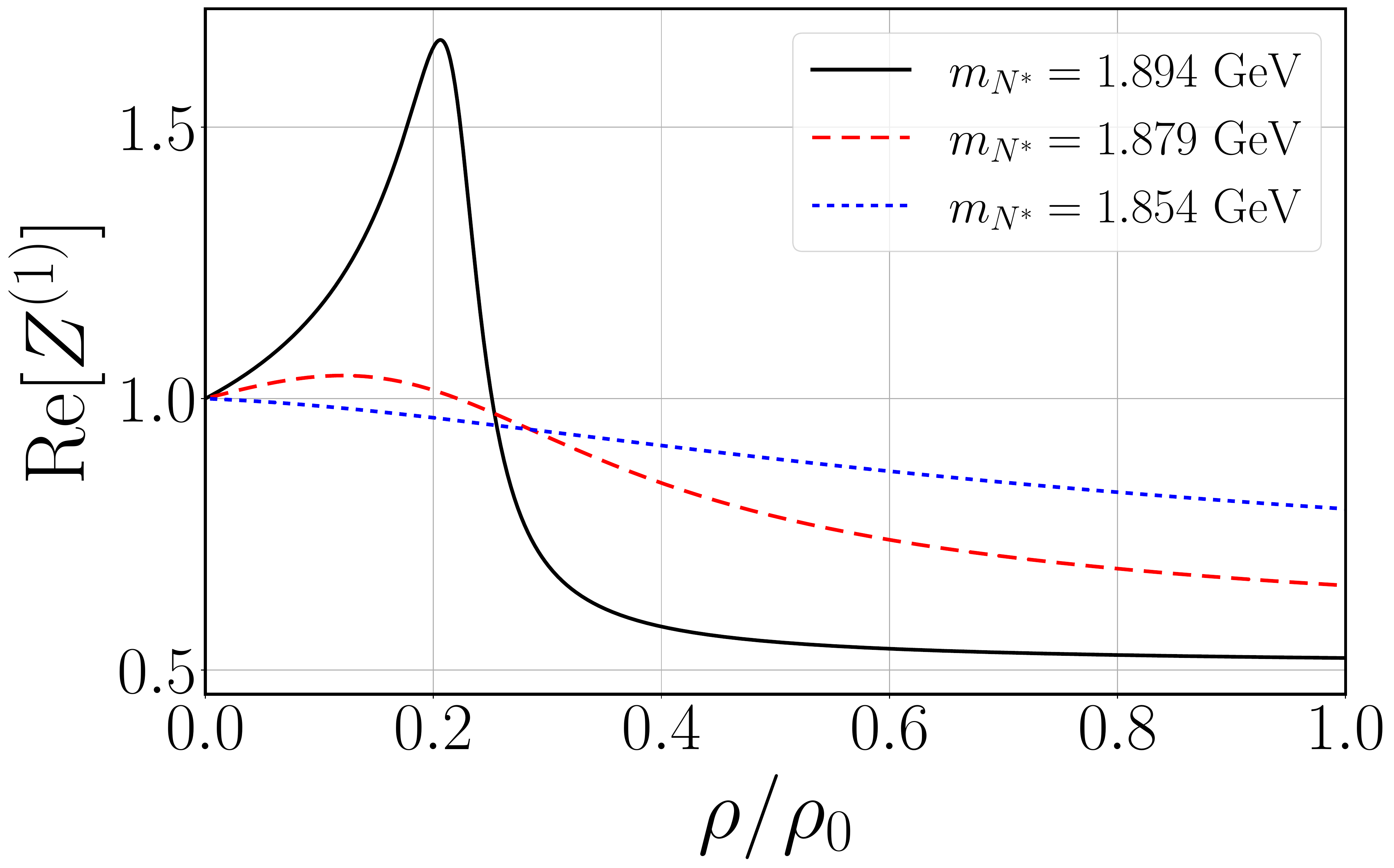}
 \caption{Real parts of the wave function renormalizations $\re[Z^{(1)}]$ calculated by the $N(1895)$-dominance model with the resonance mass $m_{N^*}=1.894$, $1.879$, and $1.854~\gev$.}
 \label{fig:residue2_resonance_comparision_rho_1}
\end{figure}

Here, we make a comment on the peculiar density dependence of the wave function renormalization seen in Fig.~\ref{fig:res_mw_res_2}.
As pointed out in Ref.~\cite{Nagahiro:2013hba}, the characteristic behavior of the residua of the poles appears when two poles approach each other.
In Fig.~\ref{fig:poletrajectory1}  we show the trajectory of the pole positions associated with the $\eta'$ and $N(1895)$-hole modes in the complex energy plane\footnote{The trajectory of $(\omega_R,-\Gamma_*/2)$ is plotted in the figure with varying the nuclear density $\rho$.}.
Two poles get close to each other when $\rho=0.2\rho_0$ to $0.3\rho_0$. In this density the wave function renormalizations, $Z^{(1)}$ and $Z^{(2)}$, have the peculiar behavior as shown in  Fig.~\ref{fig:res_mw_res_2}.
The density where two poles get close to each other can be estimated by the exceptional point discussed in Refs.~\cite{PhysRevE.61.929,Nawa:2011pz}. The exceptional point is a complex solution where these two solutions coincide. In the present case, the parameter governing the mixing of two modes, which is $\lambda$ of the linear-$\lambda$ model in Ref.~\cite{Nawa:2011pz}, is the nuclear density~$\rho$. The pole positions are determined by Eq.~\eqref{eq:pole_resmodel} with $\bp= \bm 0$, where the minor cross term is ignored for simplicity. The condition for Eq.~\eqref{eq:pole_resmodel} to have an equal root $\rho_{\rm EX}$ for the complex density is found in the linear approximation as
\begin{align}
 \frac{g_{\eta'N}^2\rho_{\rm EX}}{2m_{\eta'}}=&\frac{-1}{4}\left(m_{\eta'}+m_N-m_{N^*}+\frac{i}{2}\Gamma_{N^*}\right)^2.\label{eq:rhoex}
\end{align}
With the mass and coupling constant for $N(1895)$ given in Sec.~\ref{subsec:Nstdominance}, one finds $\rho_{\rm EX}=(0.22-i0.03)\rho_0$.
Equation~\eqref{eq:rhoex} tells us that the position of the peak in $\re[Z^{(i)}]$ is dependent on the resonance mass parameter $m_{N^*}$ and suggests that the real part of $\rho_{\rm EX}$ becomes smaller when the resonance position is away from the $\eta'N$ threshold. Thus, the characteristic behavior of $Z^{(i)}$ seen in Fig.~\ref{fig:res_mw_res_2} may get less significant for $\re[\rho_{\rm EX}]<0$.
To examine this situation, in Fig.~\ref{fig:residue2_resonance_comparision_rho_1} we show the real parts of $Z^{(1)}$ as functions of $\rho/\rho_0$ for $m_{N^*}=1.879$ and $1.854~\gev$ together with the original $m_{N^*}=1.894\ \gev$. 
From Eq.~\eqref{eq:rhoex}, the exceptional points 
for $m_{N^*}=1.879~\gev$ and $m_{N^*}=1.854~\gev$ are found to 
$\rho_{\rm EX}=(0.17-i0.22)\rho_0$ and $(-0.10-i0.53)\rho_0$,
respectively. For $m_{N^*}=1.879~\gev$, a small bump appears in $\re[Z^{(1)}]$ around the density of the real part of $\rho_{\rm EX}$, while the exceptional point for $m_{N^*}=1.854~\gev$ has a negative real part and there is no significant structure in the real part of the wave function renormalization $\re[Z^{(1)}]$. In Ref.~\cite{Nawa:2011pz}, the position of the exceptional point is discussed in the context of the nature transition. Here, we note that the nature transition does not occur in this $N(1895)$-dominance model since no level crossing takes place for the $\eta'$ and $N(1895)$-hole modes in the nuclear medium. 

\begin{figure}[t]
 \centering
 \subfigure[]{\includegraphics[width=0.45\textwidth]{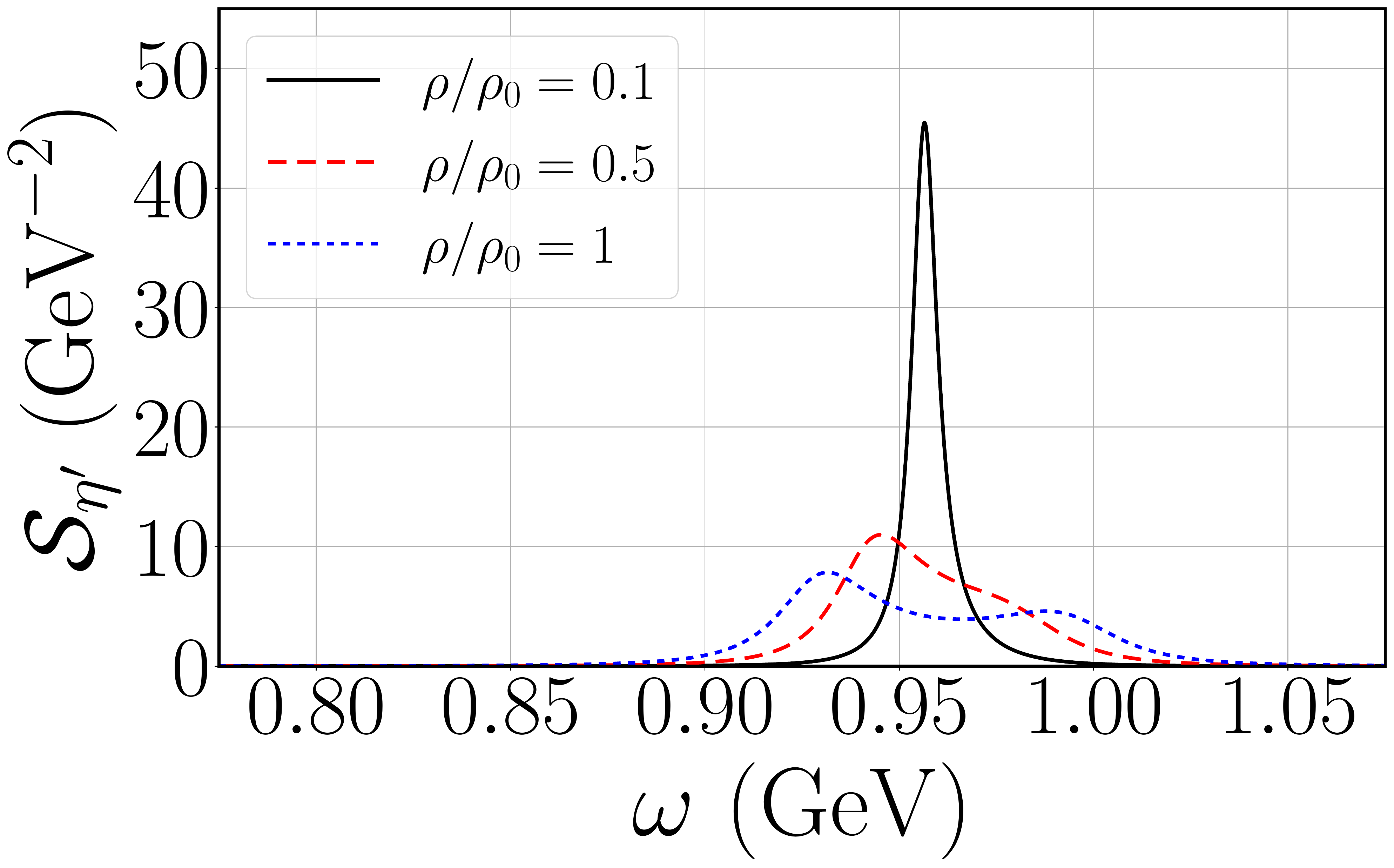}\label{fig:spec_res_2}}
 \subfigure[]{\includegraphics[width=0.45\textwidth]{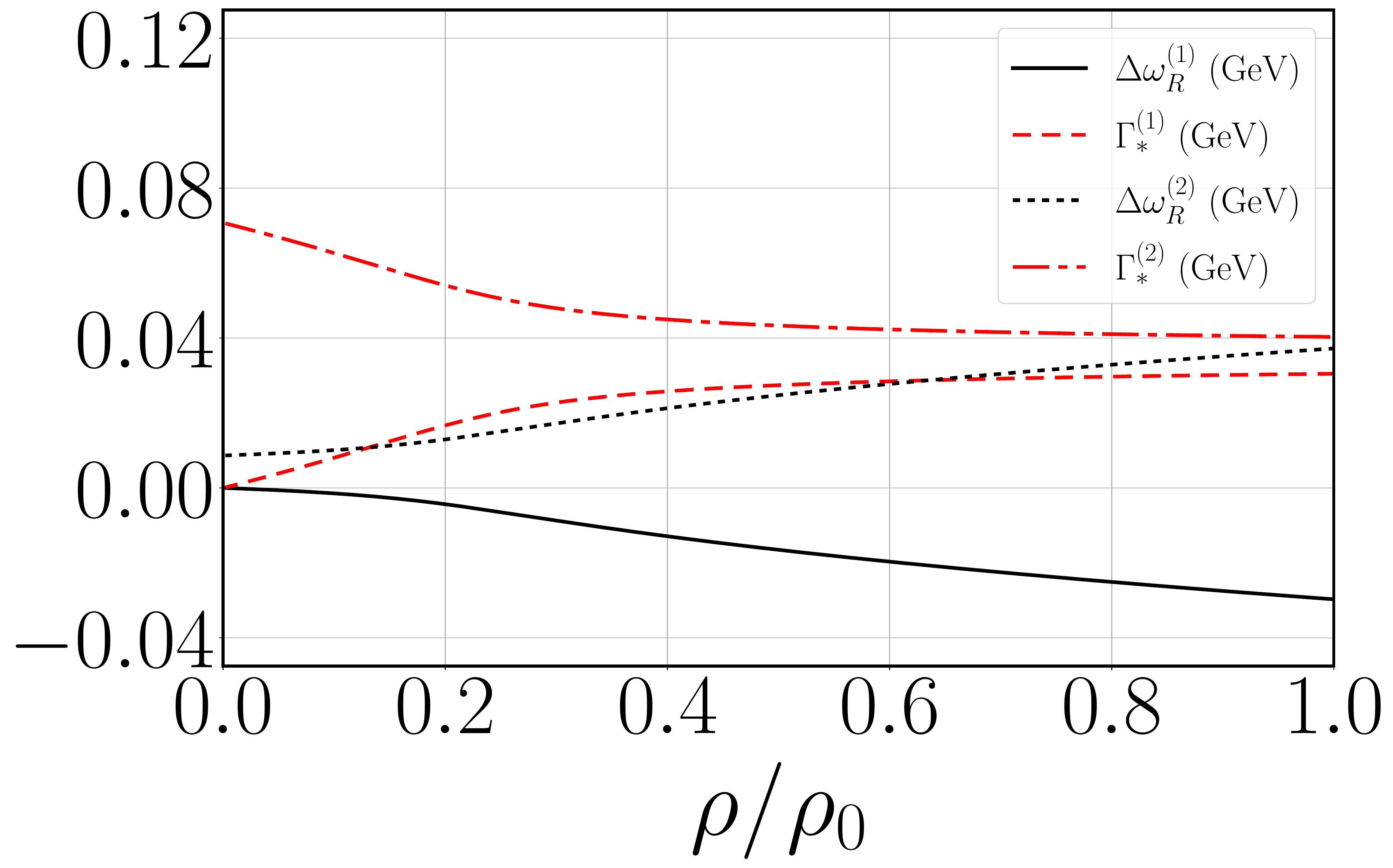}\label{fig:scat_mw_res_2}}
 \subfigure[]{\includegraphics[width=0.45\textwidth]{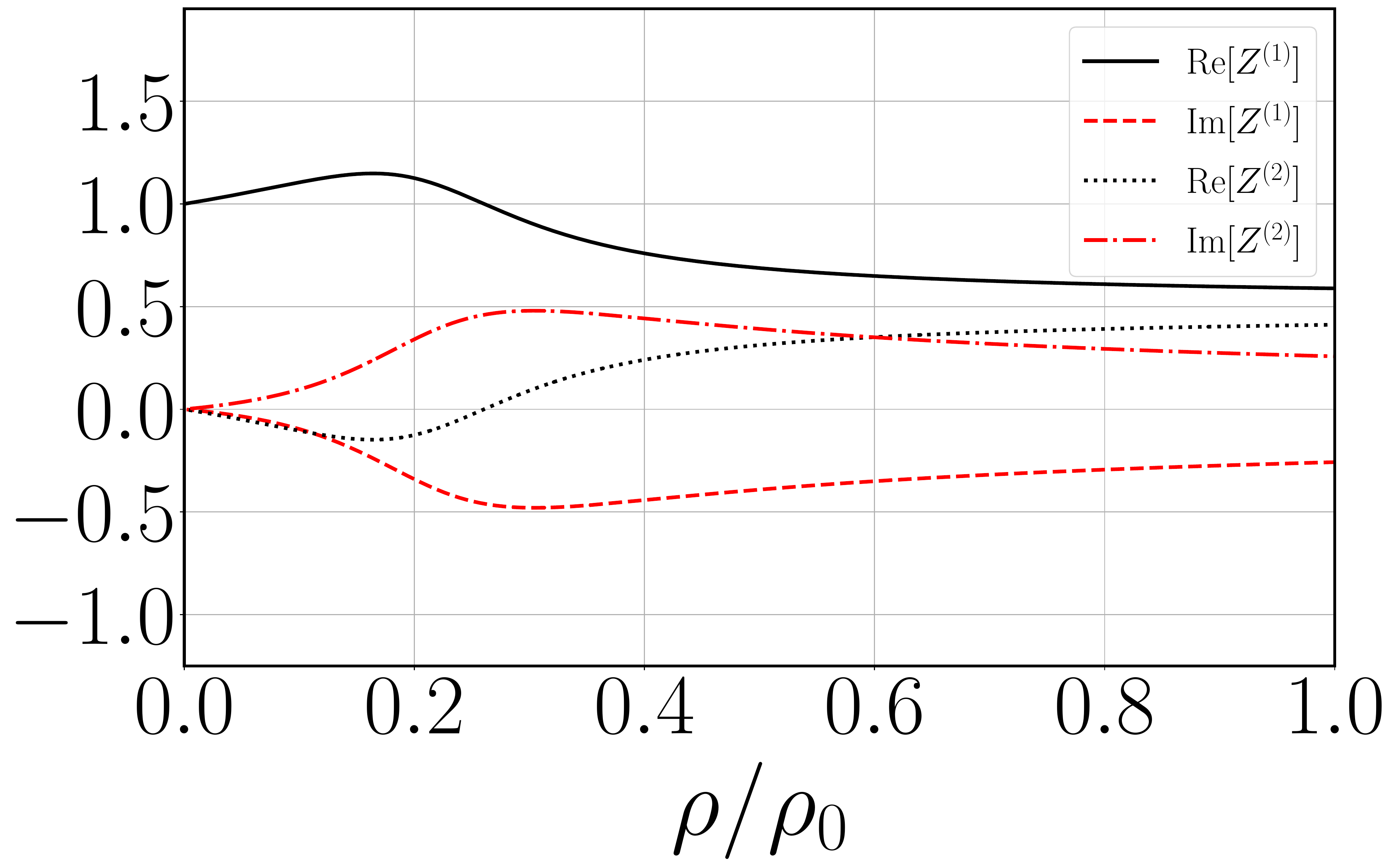}\label{fig:res_res_2}}
 \caption{
Same as Fig.~\ref{fig:spec_2} but for $m_{N^*}=1.906~\gev$
in the $N(1895)$ dominance model.
}
 \label{fig:spec_3}
\end{figure}

As mentioned in Sec.~\ref{sec:preliminaries}, the mass of $N(1895)$ can be above the $\eta'N$ threshold within the uncertainties reported by the Review of Particle Physics~\cite{ParticleDataGroup:2022pth}.
In Fig.~\ref{fig:spec_3}, we show 
the spectral function, the mass modification, the in-medium width, and the wave function renormalization
calculated in the $N(1895)$-dominance model with the resonance mass $m_{N^*}=1.906~\gev$ as suggested by Ref.~\cite{Anisovich:2017bsk}, which is located
above the $\eta'N$ threshold.
The other parameters are not changed.
As in the case with the resonance mass lower than the $\eta'N$ threshold, two peaks associated with the $\eta'$ and $N(1895)$-hole modes appear in the spectral function.
Because Fig.~\ref{fig:scat_mw_res_2} shows that the lower pole $\omega_P^{(1)}$ is connected to $\omega=m_{\eta'}$ at $\rho=0$, we identify this pole to be the $\eta'$ mode and the higher one $\omega_P^{(2)}$ the $N(1895)$-hole mode.
This order is opposite to the case with $m_{N^*}$ smaller than the $\eta'N$ threshold discussed above.
Nevertheless, the density dependences of the pole position and wave function renormalization are similar; the spectral function has one peak in lower densities, and in higher density two peaks appear and the real parts of the wave function renormalization  $Z^{(i)}$ approach $0.5$.
The characteristic peak structure in the wave function renormalization is also found in Fig.~\ref{fig:res_res_2}.
The exceptional point~\eqref{eq:rhoex} with $m_{N^*}=1.906~\gev$ is $\rho_{\rm EX}/\rho_0=0.21+i0.12$, whose real part is close to the density of the peak position in $\re[Z^{(i)}]$.
Thus, we find that the characteristic feature of the in-medium $\eta'$ spectral function in the $N(1895)$-dominance model, that is, the emergence of the two poles and the peculiar density dependences of the wave function renormalizations, are not changed even if the mass parameter of the resonance $m_{N^*}$ is 
larger than the $\eta'N$ threshold energy, 
although the order of the modes are flipped compared with the case of the resonance mass $m_{N^*}<m_{\eta'}+m_N$ as long as the resonance is located close to the threshold. 

\begin{figure}[t]
 \subfigure[]{\includegraphics[width=0.45\textwidth]{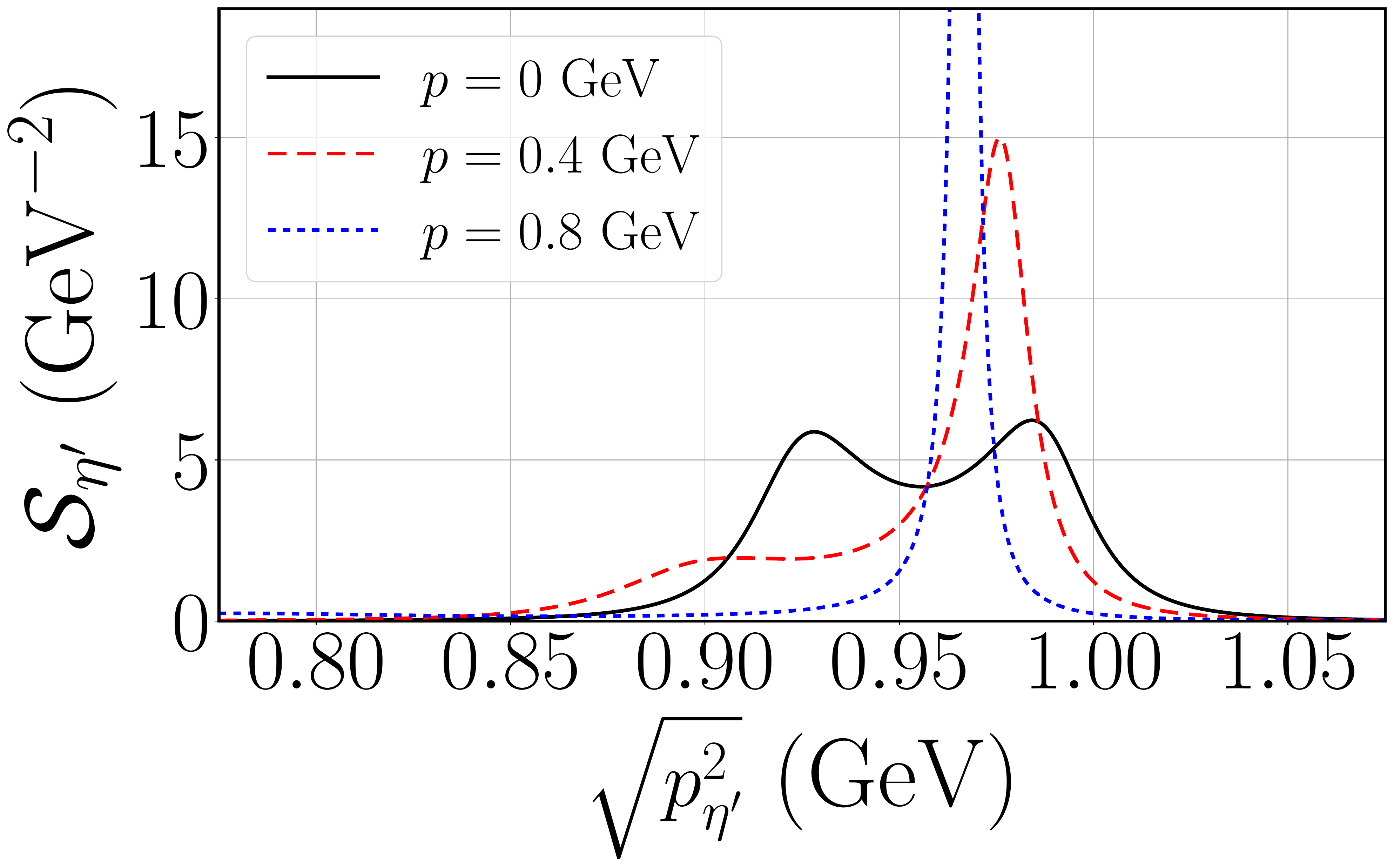}\label{fig:spec_momdep_1b}}
 \hspace{5mm}
 \subfigure[]{\includegraphics[width=0.45\textwidth]{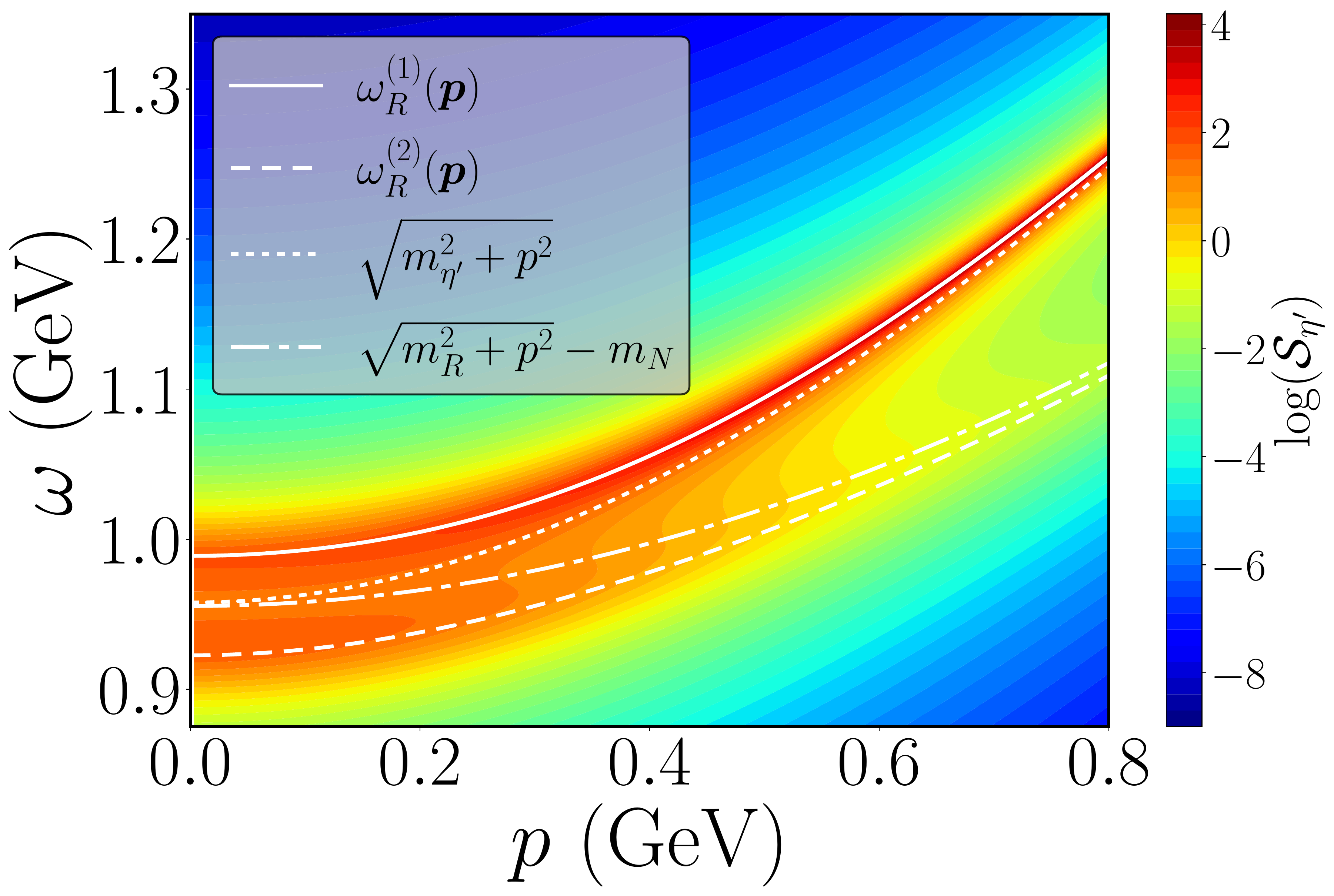}\label{fig:plot2db}}
 \caption{
 Same as Fig.~\ref{fig:spec_momdep_1} but for the $N(1895)$ dominance model.
The solid and dotted lines are the in-medium dispersion relation $\omega_R^{(1)}(\bp)$ for pole 1 and the in-vacuum $\eta'$ dispersion $\omega = \sqrt{m_{\eta'}^{2}+ \bp^{2}}$, while the dashed and the dash-dotted lines denote the in-medium dispersion relation $\omega_R^{(2)}(\bp)$ for pole 2 and the in-vacuum $N^{*}$ dispersion relation $\omega=\sqrt{m_{N^*}^2+p^2}-m_N$.
}  
 \label{fig:spec_momdep_2}
\end{figure}

In Fig.~\ref{fig:spec_momdep_2}, we show the momentum dependence of the spectral function for $\rho=\rho_{0}$ calculated with $m_{N^{*}}=1.8944~\gev$.
As discussed above, the spectral function in the $N(1895)$ dominance model has two peaks at $\rho=\rho_{0}$ for the $\eta'$ meson at rest. Figure~\ref{fig:spec_momdep_1b} shows that the height of one of the two peaks gets smaller when we turn on the $\eta'$ momentum and that for $p=0.8\ \gev$ the spectral function comes to have only one single peak. 
In Fig.~\ref{fig:plot2db}, we show the contour plot of the logarithm of the $\eta'$ spectral function $\log\left(\mS_{\eta'}\right)$ for $\rho=\rho_{0}$. In the plot, $\omega_R(\bp)$'s of the $\eta'$ and $N(1895)$-hole modes are shown by the solid and dashed lines, respectively, and we find that the $\eta'$ mode gets close to the in-vacuum $\eta'$ energy (dotted line), $\omega=\sqrt{m_{\eta'}^2+p^2}$, while the $N(1895)$-hole mode approaches the dash-dotted line, $\omega=\sqrt{m_{N^*}^2+p^2}-m_N$, which is a solution of Eq.~\eqref{eq:pole_resmodel} with finite $p$ obtained by ignoring the mixing term proportional to the nuclear density $\rho$.
Thus, the peak positions of the in-medium spectral function in larger $p$ region are mostly determined by the dispersion relation at $\rho=0$.
\begin{figure}[t]
 \centering
 \includegraphics[width=0.45\textwidth]{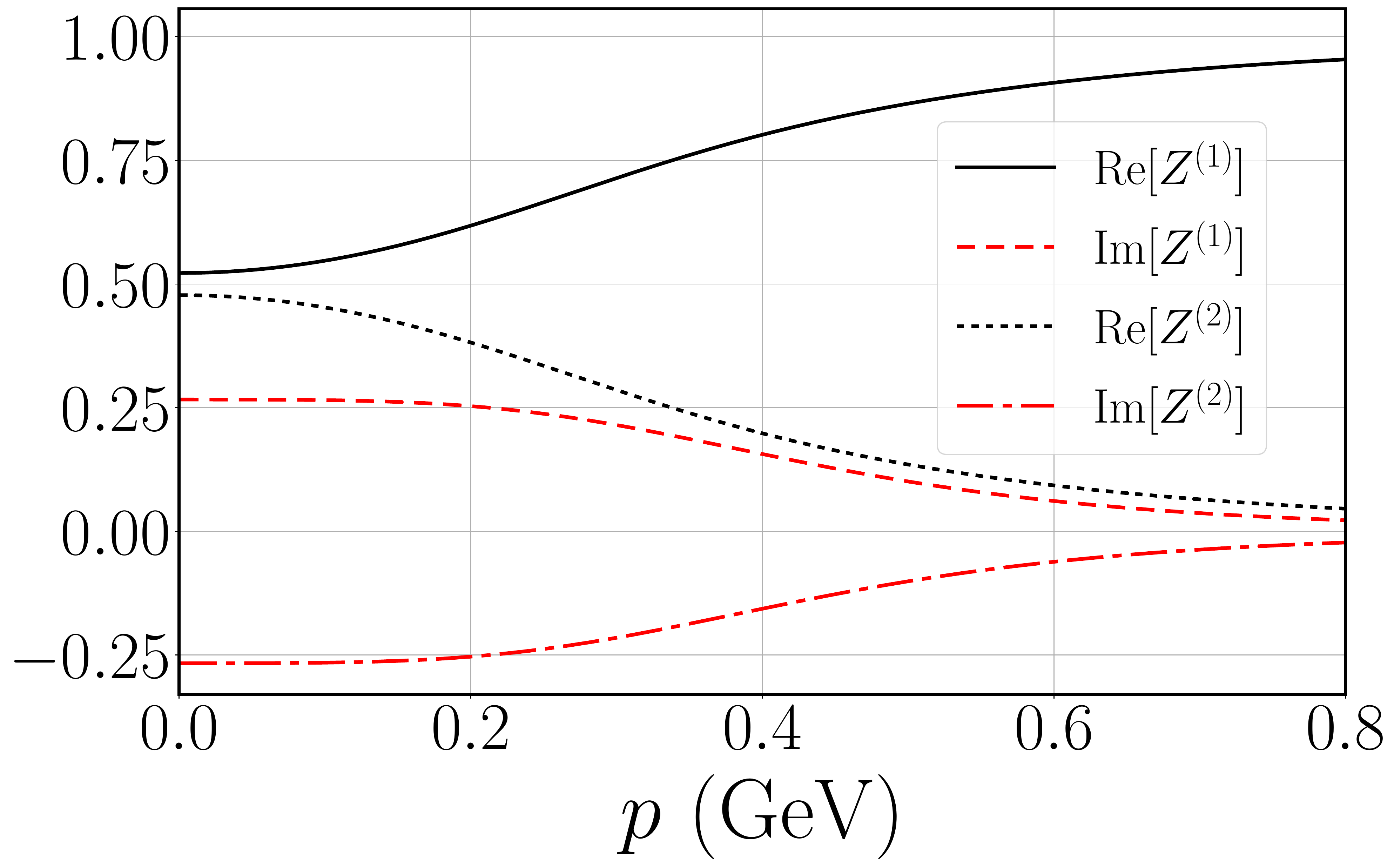}
 \caption{Plot of the real and imaginary part of $Z^{(1)}$ and $Z^{(2)}$ at $\rho=\rho_0$ in the $N(1895)$ dominance model as functions of the $\eta'$ spatial momentum $p$.
}
 \label{fig:residue_resonance_pdep}
\end{figure}
The reason that the spectral function has only one peak for higher $p$ is that the $N(1895)$-hole mode becomes irrelevant to the in-medium $\eta'$ propagation. We show the $p$ dependences of the wave function renormalizations at $\rho=\rho_{0}$ in Fig.~\ref{fig:residue_resonance_pdep}. This figure shows that the wave function renormalization of pole~2 reaches almost zero for higher momentum $p$, and the wave function renormalization of pole~1 approaches unity. Thus, the momentum dependence of the $\eta'$ spectral function can be particularly important if the $N(1895)$ resonance plays a relevant role in the in-medium $\eta'$ dynamics. 

\section{Summary}
\label{sec:summary}
In this work, we have investigated the spectral properties of the in-medium $\eta'$ mesons with zero and finite spatial momenta. The in-medium $\eta'$ self-energy is given by the $\eta'N$ scattering amplitude based on the $T\rho$ approximation. For comparison, we employ two possible models to describe the scattering amplitude;
one is the \com{coupled channels model} developed by Ref.~\cite{Bruns:2019fwi}, in which the $\eta'N$ scattering amplitude is constructed based on the chiral $U(3)$ effective Lagrangian by fixing the model parameters with the hadronic scattering data.
The other is the $N(1895)$-dominance model, in which the $\eta'N$ scattering amplitude is described by the $N(1895)$ resonance in the intermediate state. This resonance is expected to have
a sizable coupling strength to the $\eta'N$ channel from the analyses of the $\eta$ and $\eta'$ photoproduction data~\cite{Tiator:2018heh}.
With a given $\eta'N$ scattering amplitude, we investigate the in-medium $\eta'$ spectral function evaluated with the $T\rho$ approximation to focus on the qualitative feature.
In the \com{coupled channels model}, the spectral function is modified by the moderate strength of the $\eta'N$ interaction.
On the other hand, in the $N(1895)$-dominance model, the $N(1895)$ resonance introduces  energy dependence in the $\eta'$ self-energy and it causes the modification of the $\eta'$ properties in the nuclear medium.

In the \com{coupled channels model}, the spectral function has a peak and its position moves to higher energies when the density increases. The magnitude of the shift at $\rho = \rho_{0}$ is about 30 MeV, which is comparable with the size of the peak width. The width of the peak in the spectral function corresponds to the nuclear absorption of the $\eta'$ meson. 
The direction of the peak position shift reflects the repulsive nature of the $\eta'N$ interaction as expected from the negative real part of the scattering length.
The wave function renormalization, which is the residue of the pole of the in-medium propagator, is tied to the height
of the peak in the in-medium $\eta'$ spectral function. In the \com{coupled channels model} the wave function renormalization is not modified so strongly by the nuclear medium effect.
The effect of the finite spatial momentum of the $\eta'$ meson is investigated as well.
With the $\eta'$ spatial momentum $p=0.8~\gev$, the peak position is shifted about $10~\mev$ lower at $\rho=\rho_0$ compared with that with $p=0$. Although the modification is moderate, the effect can be significant.

In the $N(1895)$-dominance model, the spectral function possesses two peaks in higher densities. These peaks originate from the $\eta'$ and $N(1895)$-hole modes.
These two modes repel each other in the nuclear medium. 
Thus, if the $N(1895)$-hole mode is located below the $\eta'$ mode in vacuum, the $N(1895)$-hole mode goes down  
and the $\eta'$ mode goes up energetically as the density increases. 
Furthermore, we have found a peculiar density dependence of the wave function renormalization in the $N(1895)$-dominance model, which happens in association with the pole movement in the complex plane as the density changes.
With finite spatial momentum, the spectral function is drastically changed;
the $N(1895)$-hole mode decoupls from the $\eta'$ mode when the $\eta'$ momentum is turned on, and the spectral function approaches the one at $\rho=0$ for larger $\eta'$ momentum.

Although the difference of the spectral functions given by two models is evident, 
the shift of the real part of the pole position and 
the width \com{of the peak}
are \com{expected to be} a few tens of $\mev$ at the normal nuclear density in both models. 
The spectral function also approaches the one in free space with the increase of
the spatial $\eta'$ momentum in both models.
From the calculations done in this work, it is found that the energy and spatial-momentum dependence of the spectral function can be largely different depending on the details of the $\eta'N$ scattering process. Thus, 
the clarification of the interaction mechanism of $\eta'$ and nucleon is an important piece to understand the in-medium properties of the $\eta'$ meson and the results of experiments
of the $\eta'$-nucleus system.

\acknowledgements
We would like to thank Prof.~Naruki for her encouragement to initiate this work and
Prof.~Nagahiro for discussion on the $\eta' N$ scattering. 
The work of D.J. was partly supported by Grants-in-Aid for Scientific Research from JSPS (21K03530).

\bibliographystyle{apsrev4-2}
\bibliography{biblio}

\end{document}